\documentclass[journal]{IEEEtran}

\usepackage{cite}

\ifCLASSINFOpdf
   \usepackage{graphicx}
\else
\fi

\usepackage{amsmath}

\usepackage{url}
\RequirePackage{doi}

\usepackage{amsfonts,amssymb}
\usepackage{nccmath}
\usepackage{bm}
\usepackage{listings}
\usepackage{float}
\usepackage{comment}
\usepackage{algorithm}
\usepackage{algorithmic}
\usepackage{varwidth}
\usepackage{enumitem}
\usepackage{diagbox}
\usepackage{subfigure}
\usepackage{dsfont}
\usepackage{caption}

\usepackage{lipsum}
\allowdisplaybreaks[1]

\newtheorem{theorem}{Theorem}
\newtheorem{corollary}{Corollary}
\newtheorem{assumption}{Assumption}
\newtheorem{lemma}{Lemma}
\newtheorem{prop}{Proposition}
\newtheorem{definition}{Definition}
\newtheorem{notations}{Notations}

\newtheorem{remark}{Remark}
\newtheorem{example}{Example}

\DeclareMathOperator*{\argmax}{arg\,max}

\graphicspath{{../}}

\IEEEoverridecommandlockouts

\title{Bounded Synthesis and Reinforcement Learning of Supervisors for Stochastic Discrete Event Systems with LTL Specifications}

\author{Ryohei~Oura,
        ~Toshimitsu~Ushio,~\IEEEmembership{Member,~IEEE,}
        and Ami~Sakakibara,~\IEEEmembership{Member,~IEEE}
\thanks{
This work was partially supported by JST-ERATO HASUO Project Grant Number JPMJER1603, Japan and JST CREST Grant Number JPMJCR2012, Japan.
}
\thanks{
The authors are with the Graduate School of Engineering Science, Osaka University, Toyonaka 560-8531, Japan (e-mail: r-oura@hopf.sys.es.osaka-u.ac.jp; ushio@sys.es.osaka-u.ac.jp; amisaka@kth.se).
}
}

\begin{document}

\maketitle

\begin{abstract}

In this paper, we consider supervisory control of stochastic discrete event systems (SDESs) under linear temporal logic specifications. Applying the bounded synthesis, we reduce the supervisor synthesis into a problem of satisfying a safety condition.   First, we consider a synthesis problem of a directed controller using the safety condition.  We assign a negative reward to the unsafe states and introduce an expected return with a state-dependent discount factor.   We compute a winning region and a directed controller with the maximum satisfaction probability using a dynamic programming method, where the expected return is used as a value function. Next, we construct a permissive supervisor via the optimal value function. We show that the supervisor accomplishes the maximum satisfaction probability and maximizes the reachable set within the winning region.
Finally, for an unknown SDES, we propose a two-stage model-free reinforcement learning method for efficient learning of the winning region and the directed controllers with the maximum satisfaction probability. 
We also demonstrate the effectiveness of the proposed method by simulation.

\end{abstract}


\section{Introduction}
\label{introduction}
A discrete event system (DES) is a discrete-state event-driven system whose state changes by the occurrence of events \cite{CL2009}. A DES captures the characteristics of manufacturing systems \cite{CSX2018}, robot systems \cite{MKG2020}, and so on. The supervisory control theory was initially developed by Ramadge and Wonham \cite{RW1987}. In their framework, the DES is modeled as an automaton and the control objective is specified by a formal language. The supervisor dynamically restricts the behavior of the DES by disabling some controllable events to ensure that the controlled DES generates the specified language.
Infinite games such as mean-payoff and safety games have been utilized to synthesize a supervisor that accomplishes specified objectives. In \cite{JYL2019, JYL2021, PUK2015}, they formulated some infinite horizon objectives as mean-payoff games. A safety game was used to symbolically synthesize a permissive supervisor for timed DESs in \cite{AMP1994, AMPS1998}.

In general, more than one event are enabled at some states in the DES. Then, the nondeterminism of their occurrences is caused by its unmodeled internal structure and is often modeled by a  stochastic process quantitatively. Various stochastic discrete event system (SDES) models have been advocated. Particularly, from the perspective of supervisory control, SDESs are modeled by  probabilistic automata, probabilistic languages, and so on \cite{LW1993, PPL2009, KG2001, PL2011, DYQ2019}. In \cite{LW1993, PPL2009}, they modeled the SDES as a probabilistic automaton and initially investigated \textit{the probabilistic supervisory control problem}. They provided a necessary and sufficient condition for the existence of a probabilistic supervisor by which the controlled SDES satisfies a given probabilistic specification. They also developed a concrete algorithm for the synthesis of a probabilistic supervisor that achieves the probabilistic specification. In \cite{KG2001}, they used a probabilistic language for modeling the SDES and considered a \textit{range} control problem. The synthesized probabilistic supervisor restricts the SDES so that the generated probabilistic language lies between the lower and upper bound constraint.
In \cite{PL2011}, they addressed the case where the language specification is not achievable on the SDES and investigated an optimal supervisory control. The optimal supervisor is synthesized by minimizing the \textit{pseudometric} between the unachievable specification and its achievable approximation. Recently, in \cite{DYQ2019}, they consider the probabilistic supervisory control for the SDES under partial observation. They defined the notions of probabilistic controllability and observability and provided the polynomial verification algorithm for the notions. They demonstrated a necessary and sufficient condition of the existence of probabilistic supervisors. Besides, the optimal control problem was addressed for the case where the specification is not achievable.

In the above literature, they specified some desired properties by formal languages. However, in general, it is difficult to convert the desired property into a formal language precisely. To overcome this issue, temporal logic has been leveraged \cite{TW1986, SD1996, JSKR2006, S2020}. It has been developed in computer engineering fields as a formalism of formal specifications \cite{BK2008, CGKPV2018}. An advantage of temporal logic is the resemblance to natural languages and thus it has been widely used in several engineering fields. Particularly, complicated mission or behavior in controlled systems such as robot motion planning can be specified by temporal logic precisely and many synthesis methods of a controller or a planner that satisfy the specifications have been proposed \cite{WET2015, GZ2018, DLB2014, CXT2020}. Linear temporal logic (LTL) is often used as a specification language due to its rich expressivity. It can describe many important $\omega$-regular properties such as liveness, safety, and persistence \cite{BK2008}. It is known that any LTL formula can be converted into an $\omega$-automaton with the B\"uchi or the Rabin acceptance condition \cite{BK2008, BYG2017}.
Recently, the \textit{bounded synthesis} approach for LTL specifications was proposed \cite{SF2007,FJR2011}. Intuitively, its main idea is converting the LTL formula into a safety automaton such that all words recognized by it satisfy the LTL formula. The procedure of obtaining the safety automaton is summarized as follows. First, a given LTL specification is converted into a (universal) co-B\"uchi automaton (cBA). Next, for a non-negative integer $K$, the cBA is restricted to a $K$-co-B\"uchi automaton ($K$cBA) that rejects any word visiting the set of accepting states more than $K$ times. Finally, the $K$cBA is determinized by a usual subset construction with counters, resulting in a safety automaton. The approach has some advantages over the other automata-theoretic synthesis methods. For example, the obtained automaton is deterministic and has a smaller state space than the corresponding Rabin automaton in general. Moreover, safety automata are tractable and suitable for the synthesis of a permissive supervisor.

On the other hand, the model uncertainty is an important issue. Recently, reinforcement learning (RL) \cite{SB2018} has been paid much attention to as a useful approach to controller synthesis problems for unknown controlled systems. In general, an RL method learns an optimal control policy by trial and error on a controlled stochastic system such as a Markov decision process (MDP). In \cite{UY2003, YU2005}, they proposed an RL-based method to learn an optimal supervisor for an SDES. However, the method directly optimizes how to choose a control pattern at each state. Hence, the search space becomes exponential for the size of the event set. Furthermore, they did not consider temporal logic specifications.

Recently, automata-guided reinforcement learning methods for the synthesis of a controller or planner to achieve a given LTL specification have been numerously proposed \cite{SKCSS2014, BWZP2020-2, HPSSTW2019, HAK2018, BWZP2020, OST2020, KS2020, HKAKPL2019, LAJM2021}. This direction was initially suggested in \cite{SKCSS2014}. In the literature, they converted the given LTL specification into a deterministic Rabin automaton (DRA) and defined a reward function based on the Rabin acceptance condition. The RL of stochastic games for LTL specifications was investigated using Rabin automata in \cite{BWZP2020-2}.

However, such RL-based methods cannot be directly applied to the problem of synthesizing supervisors for SDESs under LTL specifications. More specifically, in general, supervisors directly constructed by simply combining optimal control policies obtained from \cite{SKCSS2014, BWZP2020-2, HPSSTW2019, HAK2018, HKAKPL2019, BWZP2020, OST2020, KS2020, LAJM2021} fail to accomplish the desired performance. This is because the stochastic dynamics of the Markov chain (MC) induced by the supervisor differs from those of the MCs induced by the optimal control policies. See Example 1 in \cite{JJDTK2016} for instance.
Moreover, most existing automata-guided RL methods reduced the satisfaction problem of an LTL formula into a repeated-reachability problem using the B\"uchi or the Rabin acceptance conditions. Thus, it is hard to obtain the winning region and all winning policies. 

In this paper, we introduce a value-based method for the synthesis of a permissive supervisor for the unknown SDES constrained by LTL specifications. Our main ideas are as follows.
\begin{enumerate}
    \item Inspired by the bounded synthesis \cite{SF2007, FJR2011}, we reduce the supervisor synthesis problem into the satisfaction problem of a safety condition by converting the given LTL specification into a safety automaton. We define the reward function via the acceptance condition of the product of the SDES and the automaton. 
    \item We apply a dynamic programming method to the computation of both the winning region and an optimal directed controller \cite{HK2008, HK2008-2} with the maximum satisfaction probability, where we leverage the expected return as the value function. Using the obtained optimal state value function, we construct a permissive supervisor with the maximum satisfaction probability.
    \item We propose a two-stage model-free RL method for the case where the dynamics of the SDES is unknown. For the first stage, by RL, we estimate the exact winning region and simultaneously compute all winning directed controllers. For the second stage, we compute a directed controller that maximizes the probability of reaching the estimated winning region by relearning the state value function for the states outside the winning region. Using the learned value function, we obtain the same permissive supervisor as the case using the DP-based method with probability 1.
\end{enumerate}

The contributions of our value-based bounded synthesis and RL of supervisors are summarized as three folds.
\begin{enumerate}
    \item We reduce the synthesis of the supervisor for the SDES under the LTL specification into a value-based RL for a safety condition. So, we determine the winning region by the state-value function instead of dealing directly with the state space. Furthermore, the learning problem of a desired supervisor is decomposed into the learning optimal behaviors in and outside the winning region.
    \item We leverage the Safraless procedure \cite{FJR2011}. So, the proposed method tends to make smaller state spaces than those of existing automata-guided RL methods.
    \item The proposed method can make the trade-off between the size of state space of the automaton and conservativeness for the satisfaction of the LTL specification. Moreover, the synthesized supervisor is guaranteed to be an optimal one under mild conditions.
\end{enumerate}

The rest of the paper is organized as follows. Section \ref{preliminaries} reviews an MDP, LTL, and automata. Section \ref{SDES} formulates a supervisory control problem. Section \ref{BSS} proposes a synthesis method based on the bounded synthesis and dynamic programming for directed controllers. Section \ref{MPSC} proposes the construction of permissive supervisors via the optimal value function. Section \ref{2stageRL} proposes a two-stage reinforcement learning method. Section \ref{EX} gives a numerical example to demonstrate the effectiveness of our proposed method.
\section{Preliminaries}
\label{preliminaries}

\begin{notations}
$\mathbb{N}$ is the set of positive integers. $\mathbb{N}_0$ is the set of non-negative integers. $\mathbb{R}$ is the set of real numbers. $\mathbb{R}_{\geq 0}$ is the set of non-negative real numbers. We denote the cardinality of a set $T$ by $|T|$.
Denoted by $T^\ast$ and $T^\omega$ are the sets of finite and infinite sequences obtained from a finite set $T$, respectively. For sets $L_1 \subseteq T^\ast$ and $L_2 \subseteq T^\ast \cup T^\omega$, we denote by $L_1L_2$ the concatenation of $L_1$ and $L_2$. We denote the empty string by $\varepsilon$.
\end{notations}

\subsection{Labeled Stochastic Discrete Event systems}

We define a labeled stochastic discrete event system (SDES) as a tuple $D$ = $(S, E, P_T, P_E, s^I, AP, L)$, where S is the finite set of states; $E = E_c \cup E_{uc}$ is the finite set of events partitioned into the controllable event set $E_c$ and the uncontrollable event set $E_{uc}$; $P_T:S \times S \times E \rightarrow [0,1]$ is the transition probability; $P_E : E \times S \times 2^E \rightarrow [0,1]$ is the event occurrence probability; $s^I \in S$ is the initial state; $AP$ is the finite set of atomic propositions; and $L : S \to 2^{AP}$ is the labeling function that assigns a set of atomic propositions to each state $s \in S$.

We assume that, for any state $s \in S$ and any event $e \in E$, $\sum_{s' \in S} P_T(s'|s,e) \in \{0,1\}$. Then it is said that the event $e$ is enabled at the state $s$ if $\sum_{s^\prime \in S}P_T(s^\prime|s,e)=1$.

Denoted by $E(s)$ is the sets of events enabled at the state $s$.
\begin{align}
E(s)=\{ e \;|\; \sum_{s' \in S} P_T(s' | s,e)=1 \}.
\end{align}
Let $E_c(s)$ and $E_{uc}(s)$ be the sets of controllable and uncontrollable events enabled at the state $s \in S$, respectively.
\begin{align}
E_{c}(s) &= E(s) \cap E_{c}, \\
E_{uc}(s) &= E(s) \cap E_{uc}.
\end{align}
A nonempty subset $\xi$ of $E(s)$ satisfying $E_{uc}(s) \subseteq\xi$ is called a control pattern at the state $s$. Note that $E_{uc}(s)$ is a control pattern at $s$ if and only if $E_{uc}(s) \not = \emptyset$. Let $\Xi(s) \subseteq 2^E$ be  the set of control patterns at the state $s$. Let $\Xi = \bigcup_{s \in S} \Xi(s)$.
We also assume that, for any $s \in S$ and any $\xi \in \Xi(s)$,
\begin{align}
P_E(e| s, \xi) \in \left\{
\begin{array}{ll}
(0,\ 1] & \mbox{if } e \in \xi, \\
\{0 \} & \mbox{if } e \not\in \xi,
\end{array}  \right.
\end{align}
and
\begin{align}
\sum_{e \in \xi} P_E(e | s, \xi)=1.
\end{align}

The controlled transition function $P:S \times \{ (s,\xi) | s \in S, \xi \in \Xi(s) \} \to [0,1]$ is defined as follows:  for any $s, s' \in S $ and any $\xi \in \Xi(s)$,
\begin{align}
P(s' | s, \xi)=\sum_{e \in \xi}P_E(e | s, \xi)P_T(s' | s, e).
\end{align}
Note that, for any $s \in S$ and any $ \xi \in \Xi(s)$, we have
\begin{align}
\sum_{s' \in S}P(s' | s, \xi)=1.
\end{align}




Note that, in this paper, we do not choose an event but a control pattern deterministically. An event $\sigma$ is said to be control-enabled (with respect to a control pattern $\xi$) if $\sigma \in \xi$ \cite{RW1987}. A control pattern $\xi$ at the state $s$ is called a \textit{directed control pattern} if  $|\xi \cap E_c(s)|\leq 1$ \cite{HK2008, HK2008-2}. For convenience, we will write $\xi_e$ for the directed control pattern including the controllable event $e$. Let $\xi_{\emptyset}(s) = E_{uc}(s)$. Denoted by $\Xi_{dir}(s)$ is the set of directed control patterns at the state $s$. Note that $\xi_{\emptyset}(s) \in \Xi_{dir}(s)$ if and only if $E_{uc}(s)$ is nonempty. Moreover, let $\Xi_{dir} = \bigcup_{s \in S} \Xi_{dir}(s)$.
\begin{remark}
 Our definition of the SDES generalizes the definition of the probabilistic generator and the automaton in \cite{LW1993} and \cite{KG2001}, respectively. When we assume that the state transition is deterministic and each event in a given $\xi$ occurs with the probability in accordance with the proportion to their occurrence probability under the control pattern by which all events are control-enabled, our definition coincides with the defined SDESs in \cite{LW1993} and \cite{KG2001}.
\end{remark}

In the SDES $D$, an infinite path starting from a state $s_0 \in S$ is defined as a sequence $\rho\ =\ s_0e_0s_1 \ldots\ \in S (E S)^{\omega}$ such that $e_i \in E(s_i)$ and $P_T(s_{i+1}|s_i, e_i) > 0$ for any $ i \in \mathbb{N}_0$. A finite path is a finite sequence in $S ( E S)^\ast$.
For a path $\rho\ =\ s_0e_0s_1 \ldots$, we define the corresponding labeled path $L(\rho)\ =\ L(s_0)L(s_1) \ldots \in (2^{AP})^{\omega}$.
 $InfPath^{D}(s)$ (resp., $FinPath^{D}(s)$) is defined as the set of infinite (resp., finite) paths starting from $s_0=s$ in the MDP $D$ and, for simplicity, $InfPath^D(s^I)$ (resp., $FinPath^D(s^I)$) is denoted by $InfPath^D$ (resp., $FinPath^D$). For each finite path $\rho$, $last(\rho)$ denotes its last state.
 For the SDES $D$, the smallest $\sigma$-algebra over all possible infinite paths is constructed with a usual way and the unique probability measure $Pr^D$ on it is defined \cite{BK2008}. For $ P \subseteq InfPath^D$ whose all infinite paths contain a finite path $s_0 e_1 s_2 \ldots s_{\ell} \in FinPath^D$ as their prefix, its probability is given by $Pr^D(P) = \prod_{i=0}^{\ell-1} P_T(s_{i+1}|s_i,e_i)P_E(e_i|s_i,E(s_i))$.

\begin{figure}[htbp]
  \centering
  	\includegraphics[width = 7cm]{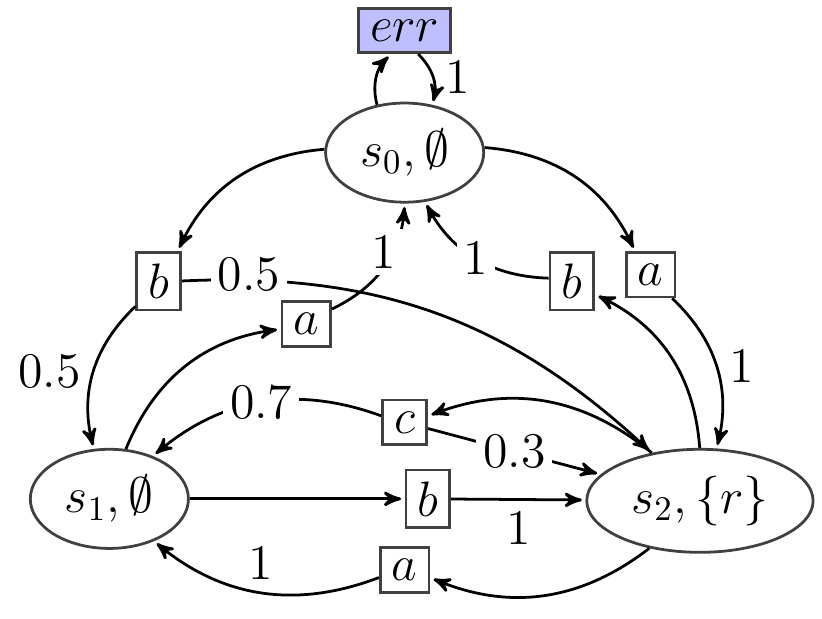}
  	\caption{An illustrative SDES with 3 states. Each value labeled with each edge between a state and an event is the transition probability. The event $err$ is uncontrollable and the others are controllable. $s_2$ is labelled by $\{r\}$.}
  	\label{SDES_S3}
  \end{figure}

\begin{figure}[htbp]
  \centering
  	\includegraphics[width = 8cm]{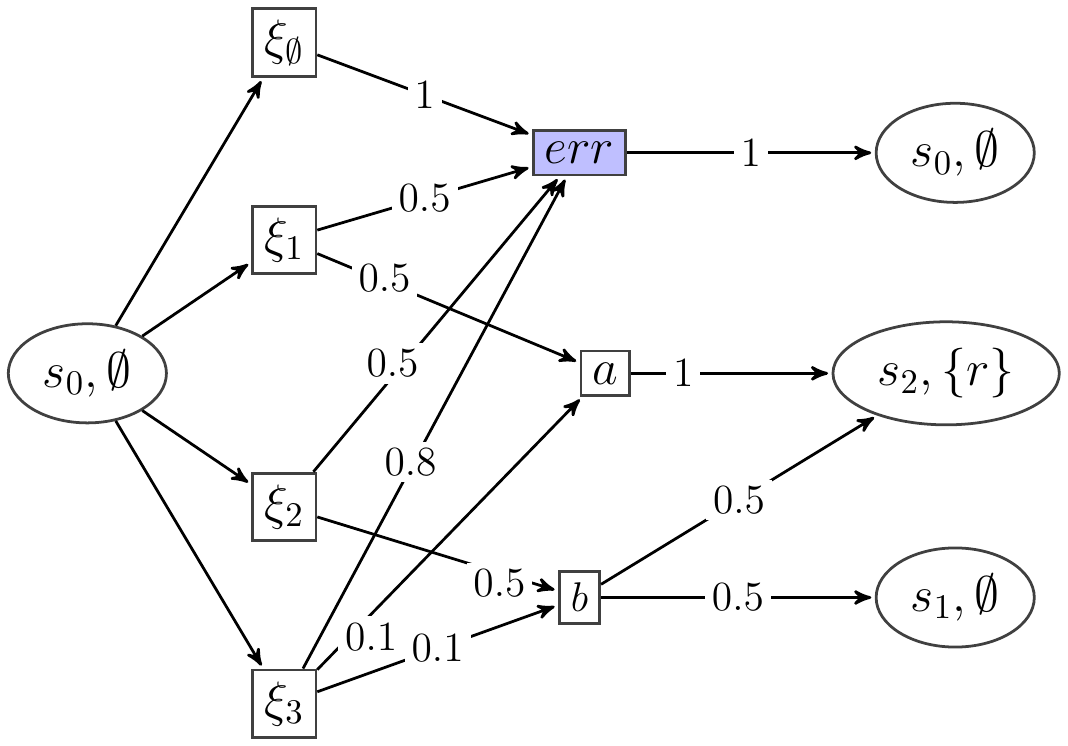}
  	\caption{A part of the SDES. Each value labeled with each edge between a control pattern and an event is the event occurrence probability.}
  	\label{SDES_event}
  \end{figure}

We show an illustrative example of the transition properties of an SDES $(S, E = E_c \cup E_{uc}, P_T, P_E, s^I, AP, L)$ in the Fig.\! \ref{SDES_S3}, where $S=\{s_0, s_1,s_2\}$, $s^I=s_1$, $E_c=\{a, b, c\}$, $E_{uc}=\{ err\}$, $AP=\{r\}$, and the labeling function $L$ is defined as $L(s_2)=\{ r \}$ and $L(s) = \emptyset$ for $s=s_0,s_1$. Only the transitions from $s_0$ and $s_2$ with the event $b$ and $c$ are probabilistic. Each transition from $s_0$ to $s_1$ and $s_2$ with $b$ occurs with probability 0.5. The transition from $s_2$ to $s_1$ with $c$ occurs with probability 0.7 and, with probability 0.3, the state stays $s_2$. Other than the state $s_0$, the event occurrence probabilities is equal to the inverse of the number of control-enabled events. That is, for each state $s \in S \setminus \{ s_0 \}$, 
$
P_E(e | s,\xi)=\frac{1}{|\xi|} \mbox{ for each }\xi \in \Xi(s)\mbox{ and any } e \in \xi.
$
We show a part of the SDES including the event occurrence probability at $s_0$ in Fig.\ \ref{SDES_event}. Let $\xi_\emptyset = \{ err \}$, $\xi_1 = \{a, err \}$, $\xi_2 = \{b, err \}$, and $\xi_3 = \{ a, b, err \}$. The values at the edges between control patterns and events are the event occurrence probabilities.
  
\subsection{Linear Temporal Logic and Automata}

We use linear temporal logic (LTL) formulas to describe various temporal constraints or properties specified to the SDES.
LTL formulas are constructed from a set of atomic propositions, Boolean operators, and temporal operators. We use the standard notations for the Boolean operators: $\top$ (true), $\neg$ (negation), and $\land$ (conjunction).
LTL formulas over a set of atomic propositions $AP$ are defined as
\begin{align*}
  \varphi ::=\top\ |\ \alpha \in AP\ |\ \varphi_1 \land \varphi_2\ |\ \neg \varphi\ |\ \text{{\bf X}} \varphi\ |\ \varphi_1 \text{{\bf U}} \varphi_2,
\end{align*}
where $\varphi$, $\varphi_1$, and $\varphi_2$ are LTL formulas.
Additional Boolean operators are defined as $\perp := \neg \top $, $\varphi_1 \lor \varphi_2 := \neg(\neg \varphi_1 \land \neg \varphi)$, and $\varphi_1 \Rightarrow \varphi_2 := \neg \varphi_1 \lor \varphi_2$.
The operators {\bf X} and {\bf U} are called ``next" and ``until", respectively.

Let $\rho[i:]$ be the $i$-th suffix $\rho[i:]=s_ie_is_{i+1} \ldots $ and $\rho[i]$ be the $i$-th state $\rho[i] = s_i$.

\begin{definition}
	For an LTL formula $\varphi$ 
	and an infinite path $\rho = s_0a_0s_1 \ldots$ of a DES $ D $ with $ s_0 \in S $, the satisfaction relation $D,\rho \models \varphi$ is recursively defined as follows.
	\begin{alignat}{2}
	& D, \rho \models \top,\nonumber \\
	& D, \rho \models \alpha \in AP &&\Leftrightarrow \alpha \in L(\rho[0]),\nonumber \\
	& D, \rho \models \varphi_1 \land \varphi_2 &&\Leftrightarrow D, \rho \models \varphi_1 \land D, \rho \models \varphi_2,\nonumber \\
	& D, \rho \models \neg \varphi &&\Leftrightarrow D, \rho \not\models \varphi,\nonumber \\
	& D, \rho \models \text{{\bf X}}\varphi &&\Leftrightarrow D, \rho[1:] \models \varphi,\nonumber \\
	& D, \rho \models \varphi_1 \text{{\bf U}} \varphi_2 &&\Leftrightarrow \exists j \geq 0, \ D, \rho[j:] \models \varphi_2 \nonumber \\
	& &&\quad \land \forall i, 0\leq i < j, \ D, \rho[i:] \models \varphi_1. \nonumber
	\end{alignat}

The next operator {\bf X} requires that $\varphi$ is satisfied by the next state suffix of $\rho$. The until operator {\bf U} requires that $\varphi_1$ holds true until $\varphi_2$ becomes true over the path $\rho$. Using the operator {\bf U}, we define two temporal operators: 1) {\it eventually}, $\text{{\bf F}} \varphi := \top \text{{\bf U}} \varphi $ and 2) {\it always}, $\text{{\bf G}} \varphi := \neg \text{{\bf F}} \neg \varphi$.
In the following, we write $ \rho \models \varphi $ without referring to $ D $ for simplicity.
\end{definition}

We define an $\omega$-automaton.

\begin{definition}[$\omega$-automaton]
  An $\omega$-automaton is a tuple $A = (X,\Sigma,\delta, x^I, Acc)$, where $X$ is the finite set of states, $\Sigma$ is the input alphabet including $\varepsilon$, $\delta :  X\times \Sigma \to 2^X$ is the transition function, $x^I \in X$ is the initial state, and $Acc$ is the accepting set, namely the set of accepting states.
\end{definition}


An infinite sequence $w \in \Sigma^{\omega}$ is called a word. An infinite sequence $r = x_0\sigma_0x_1 \ldots \in X (\Sigma X)^{\omega}$ is called a run on $A$ generated by a word $w = \sigma_0 \sigma_1 \ldots \Sigma^\omega$ if $x_{i+1} \in \delta(x_i, \sigma_{i})$ for any $ i\in \mathbb{N}_0$.  For an $\omega$-automaton $A$ and a word $w$, we denote by $Runs(w; A)$ the set of runs on $A$ generated by $w$. Moreover, for a state $x$ and a run $r$, we denote by $Visits(x; r)$ the number of times $r$ visits $x$.

We define the universal co-B\"uchi and the universal $K$-co-B\"uchi automaton.

\begin{definition}[universal co-B\"{u}chi and universal $K$-co-B\"uchi automaton]
  Let $w \in \Sigma^\omega$ be a word. An $\omega$-automaton $A$ with the following universal co-B\"uchi acceptance condition is called a universal co-B\"uchi automaton (cBA).
 \begin{itemize}
 \item \textbf{ The universal co-B\"uchi acceptance condition}: $w$ is accepted by $A$  if and only if, for any $r \in Runs(w;A)$ and any $x \in Acc$, $Visits(x; r)<\infty$.
 \end{itemize}

 For a non-negative integer $K$, an $\omega$-automaton $A$ with the following universal $K$-co-B\"uchi acceptance condition is called universal $K$-co-B\"uchi automaton ($K$cBA).
 \begin{itemize}
 \item \textbf{ The universal $K$-co-B\"uchi acceptance condition}: $w$ is accepted by $A$  if and only if, for any $r \in Runs(w;A)$, $\sum_{x \in Acc} Visits(x; r) \leq K$.
 \end{itemize}

\end{definition}
Denoted by ${\mathcal L}_c(A)$ and  ${\mathcal L}_{c,K}(A)$ are the sets of words accepted by the $\omega$-automaton with the universal co-B\"uchi and the universal $K$-co-B\"uchi acceptance condition, respectively.
Moreover, to clarify which acceptance condition is adopted, the cBA and the $K$cBA are denoted by $B$ and $(B,K)$, respectively.

Note that, for any $\omega$-automaton $A$ and any non-negative integers $K_1, K_2 \in \mathbb{N}_0$, if $K_1 \leq K_2$ then we have $\mathcal{L}_{c,K_1}(A) \subseteq \mathcal{L}_{c,K_2}(A) \subseteq \mathcal{L}_{c}(A)$ \cite{FJR2011}.


We determinize the $K$cBA by a normal subset construction with counters \cite{FJR2011}.
\begin{definition}[Determinization of $K$cBA]
For a $K$cBA $(B,K) = (X, \Sigma, \delta, x^I, Acc)$, its determinization (d$K$cBA) is a tuple $det(B,K) = (\mathcal{F}, \Sigma, \Delta, F^I, Acc_{d})$, where:

\begin{itemize}
    \item $\mathcal{F} = \{ F | F \text{ is a mapping from } X \text{ to } \{-1, \ldots, K+1\} \}$.
    \item For $F \in \mathcal{F}$ and $\sigma \in \Sigma$, $\Delta(F,\sigma) : X \to \{-1, \ldots, K+1\}$ is a mapping such that
     $\Delta(F,\sigma)(x^\prime) = \max \{ \min ( K+1, F(x) + (x^\prime \in Acc)) | x^\prime \in \delta(x, \sigma), F(x) \neq -1 \}$, where $\max \emptyset = -1$.
     
    \item $F^I : X \to \{-1, 0, 1\}$ is a mapping such that
    \begin{align*}
       F^I(x) =
        \left\{
    \begin{aligned}
      &-1 &&\text{ if } x \neq x^I, \\
      &(x \in Acc) &&\text{ otherwise, }
    \end{aligned}
    \right.
    \end{align*}
    where $(x \in Acc) = 1$ if $x$ is in $Acc$, otherwise $(x \in Acc) = 0$.

     \item $Acc_{d} = \{ F \in \mathcal{F} \;|\; \exists x \in X \text{ s.t. } F(x) > K \}$.
\end{itemize}
\end{definition}
Intuitively, the determinization is constructed by, for all $x \in X$, keeping track of the maximal number of accepting states that have been visited by runs ending up $x$ using the counters $F \in \mathcal{F}$. For each $x \in X$, the count for $x$ is set to -1 if no run starting from $x^I$ ends up in $x$. The accepting states are represented by the set of $F \in \mathcal{F}$ such that an original state $x$ has a count greater than $K$, i.e., $F(x) > K$.

We say that an $\omega$-automaton $A$ is deterministic if $| \delta(x, \sigma) | \leq 1$ for any $x \in X$ and any $\sigma \in \Sigma$. $A$ is complete if, for any $w \in \Sigma^\omega$, $ Runs(w; A) \neq \emptyset$ holds.

The following proposition is shown in \cite{FJR2011}.
\begin{prop}
\label{prop_detK}
Let $B$ be a complete cBA. The corresponding $det(B,K)$ is deterministic, complete, and $\mathcal{L}_{c,K}(B) = \mathcal{L}_{c,0}(det(B,K))$.
\end{prop}

It is known that, for any LTL formula $ \varphi $, there exists a complete cBA that accepts all words satisfying $\varphi$ \cite{FJR2011}.
In particular, we represent a complete cBA recognizing an LTL formula $\varphi$ as $B_{\varphi}$, whose input alphabet is given by $ \Sigma = 2^{AP} $.
By Proposition \ref{prop_detK}, there exists a d$K$cBA constructed as a $0$-co-B\"uchi automaton from a given LTL formula such that all words recognized by the d$K$cBA satisfy the LTL formula.

Note that, for a d$K$cBA, the size of $\mathcal{F}$ is about $K^{|Acc|}$ times larger than the set of states $X$ of the original cBA in the worst case.

\begin{definition}[Sink set]
Let $A = (X,\Sigma,\delta, x^I, Acc)$ be an $\omega$-automaton. For the subset of states $X_{sub} \subseteq X$, we say that $X_{sub}$ is a sink set if there is no outgoing transition from $X_{sub}$ to $X \setminus X_{sub}$, that is $ \delta(x, \sigma) \subseteq X_{sub}$ for any $x \in X_{sub}$ and any $\sigma \in \Sigma$.
\end{definition}

Note that the set of accepting sates $Acc_d$ can be constructed as a sink set for any d$K$cBA $det(B,K)$. This is because, for any run $r \in X (\Sigma X)^\omega$, once $r$ enters $Acc_d$, it never satisfies the acceptance condition of $det(B,K)$. Practically, $Acc_d$ can be constructed as a singleton.

\begin{figure}[hbtp]
  	\centering
  	\begin{minipage}[b]{0.45\linewidth}
  	    \centering
  		\includegraphics[height = 3.8cm]{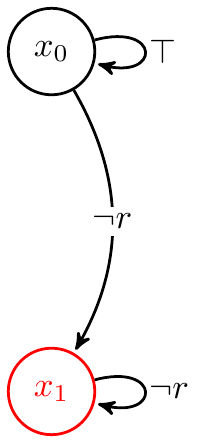}
  		\caption*{(a)}
  	\end{minipage}
  	\begin{minipage}[b]{0.45\linewidth}
  	    \centering
  		\includegraphics[height=3.8cm]{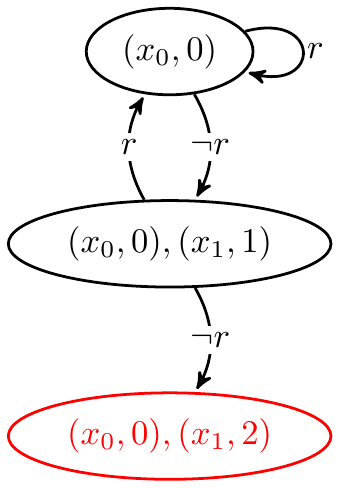}
  		\caption*{(b)}
  	\end{minipage}
  	\begin{minipage}[b]{0.45\linewidth}
  	    \centering
  		\includegraphics[height = 4.7cm]{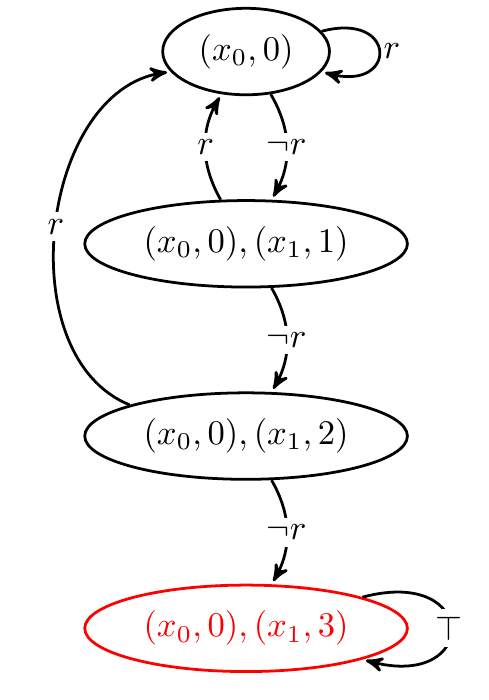}
  		\caption*{(c)}
  	\end{minipage}
  	\begin{minipage}[b]{0.45\linewidth}
  	    \centering
  		\includegraphics[height=4.7cm]{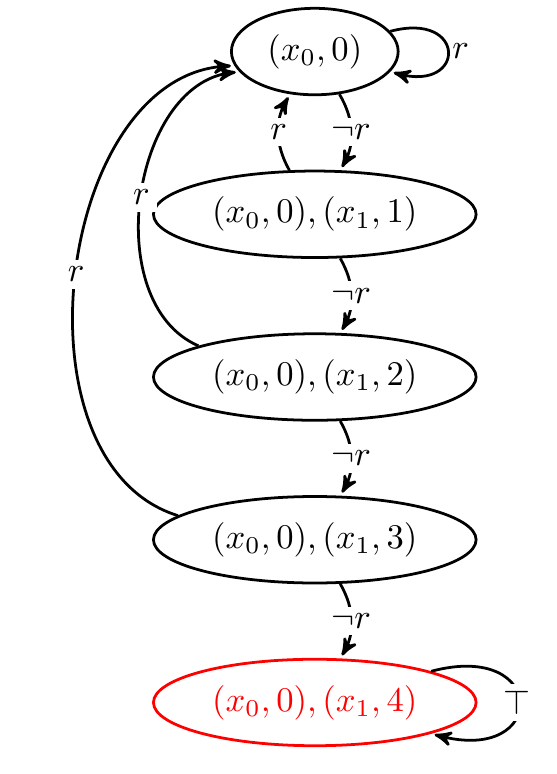}
  		\caption*{(d)}
  	\end{minipage}
  	\caption{(a) A complete cBA converted from $\varphi = \textbf{GF}r$. (b), (c), (d) Its determinization with $K=1,2,3$. Their accepting states are colored with red.}
  	\label{fig:cBA_GFa}
  \end{figure}

We consider an LTL specification $\varphi = \textbf{GF} r$ that represents the specification ``the state labeled by $\{r\}$ is visited infinitely often''. Then, shown in Fig.\ \ref{fig:cBA_GFa} are a complete cBA and its corresponding complete dKcBAs with $K = 1,2,3$ converted from $\varphi$. We represent their accepting states with red circles. Note that states of the automaton are represented by the pair of the state of the original $K$-co-B\"uchi automaton and the number counted by the counters $F$ of the d$K$cBA. We omit the states of the automaton that are not reachable from the initial state.

\subsection{Product Stochastic Discrete Event Systems}
We introduce the product of an SDES and a complete d$K$cBA.

\begin{definition}[Product SDES]
\label{product_def}
For a complete d$K$cBA $det(B,K) = (\mathcal{F}, \Sigma, \Delta, F^I, Acc_d)$ and an SDES $D = (S, E, P_T, P_E, s^I, AP, L)$, the product SDES is a tuple $D^{\otimes} = D \otimes det(B_{\varphi},K)$ $= (S^{\otimes}, E^{\otimes}, P^{\otimes}_T, P^{\otimes}_E, s^{\otimes I}, Acc^{\otimes})$, where
$S^{\otimes} = S \times \mathcal{F}$ is the finite set of states;
$E^{\otimes}=E$, $E^\otimes_c = E_c$, and $E^\otimes_{uc} = E_{uc}$; $P^{\otimes}_T : S^{\otimes} \times S^{\otimes} \times E^{\otimes} \rightarrow [0,1]$ is the transition probability defined as, for any $s^\otimes = (s,F)$ and any $e \in E^\otimes(s)$,
\begin{align}
  &P^{\otimes}_T(s^{\otimes \prime} | s^{\otimes}, e) =\left\{
  \begin{aligned}
    &P_T(s^{\prime} | s, e) &   &\text{if}\ F^{\prime} = \Delta(F, L(s)), \\
    &0 &   &\text{otherwise} ,
  \end{aligned}
  \right. 
\end{align}
where $s^{\otimes \prime} = (s^{\prime}, F^{\prime})$;
$P^{\otimes}_E : E^{\otimes} \times S^{\otimes} \times 2^{E^{\otimes}} \rightarrow [0,1]$ is the event occurrence probability defined as $P^{\otimes}_E(e | s^{\otimes}, \xi) = P_E(e | s, \xi)$ for any $s^\otimes = (s,F) \in S^\otimes$ and any $\xi \in \Xi(s)$; $s^{\otimes I} = (s^{I},\hat{F}^{I})$ is the initial states, where $\hat{F}^I = \Delta(F^I, L(s^I))$, and $Acc^{\otimes} = S \times Acc_d$.
\end{definition}
Intuitively, the product SDES is a synchronized structure between the SDES and the d$K$cBA. In other words, the product SDES represents simultaneously the transition of the SDES and the associated transition of the d$K$cBA.

Note that $Acc^\otimes$ is non-empty since $Acc_d$ is non-empty.
For simplicity, the product SDES $D^{\otimes}$ of a given SDES and a d$K$cBA converted from an LTL formula $\varphi$ will be called a product SDES associated with $\varphi$.

For each $s^\otimes = (s,F) \in S^\otimes$, let $E^\otimes(s^\otimes) = E(s) $, $E^\otimes_c(s^\otimes) = E_c(s)$, and $E^\otimes_{uc}(s^\otimes) = E_{uc}(s)$. Likewise, let $\Xi^\otimes_{dir} = \Xi_{dir}$ and  $\Xi^\otimes_{dir}(s^\otimes) = \Xi_{dir}(s) $ for each $s^\otimes = (s,F) \in S^\otimes$.
The controlled transition probability $P^\otimes : S^\otimes \times \{(s, \xi) | s \in S^\otimes, \xi \in \Xi^\otimes(s) \} \to [0,1]$ is defined as, for any $s, s' \in S^\otimes$ and any $\xi \in \Xi^\otimes(s)$, $P^\otimes(s^\prime|s,\xi) = \sum_{e \in \xi}P^\otimes_T(s^\prime|s,e) P^\otimes_E(e|s,\xi)$.
For convenience, we sometimes omit the superscripts $\otimes$.

Note that $Acc^\otimes$ has no outgoing transition to $S^\otimes \setminus Acc^\otimes$ since $Acc_d$ is a sink set. Thus, once a path of the product SDES enters $Acc^\otimes$, from then on, its suffix always stays in $Acc^\otimes$.

For any product SDES $D^\otimes$, its acceptance condition is a safety condition since it is satisfied when any path generated on $D^\otimes$ always stays in $S^\otimes \setminus Acc^\otimes$.
In the following, we call $s \in S^\otimes \setminus Acc^\otimes$ a \textit{safe state} and $s \in Acc^\otimes$ an \textit{unsafe state}.

For a subset $S_{sub}^\otimes$ of $S^\otimes$, we introduce an atomic proposition ``This state belongs to $S_{sub}^\otimes$'', which denotes $S_{sub}^\otimes$ by abuse of notation, that is, we say that a state $s\in S^{\otimes}$ satisfies $S^\otimes_{sub}$ if $s \in S^\otimes_{sub}$. Then, the acceptance condition of $D^\otimes$ is represented by
\begin{align}
\label{acc_cond_prodSDES}
\varphi_B = \textbf{G} \neg Acc^\otimes,
\end{align}
that is, an infinite path $\rho\in InfPath^{D^{\otimes}}$ is accepted by the SDES $D^{\otimes}$ if and only if $\rho\models \varphi_B$.

The \textit{determinized logical part} of $D^\otimes = (S^{\otimes}, E^{\otimes}, P^{\otimes}_T,$ $ P^{\otimes}_E, s^{\otimes I}, Acc^{\otimes})$ is a tuple $logic(D^\otimes) = (S^\otimes , E \times S^\otimes, \delta,$ $s^{\otimes I})$, where, $\delta : S^\otimes \times (E \times S^\otimes) \to S^\otimes$ is the transition function defined as, for each $(s^\otimes, (e, s^{\otimes \prime})) \in S^\otimes \times (E \times S^\otimes)$, $\delta(s^\otimes, (e,s^{\otimes \prime})) = s^{\otimes \prime}$ if $P^\otimes_T(s^{\otimes \prime} \;|\; s^\otimes ,e) > 0 $, otherwise $\delta(s^\otimes, (e,s^{\otimes \prime}))$ is undefined.

\begin{figure}[htbp]
  \centering
  	\includegraphics[width = 8.6cm]{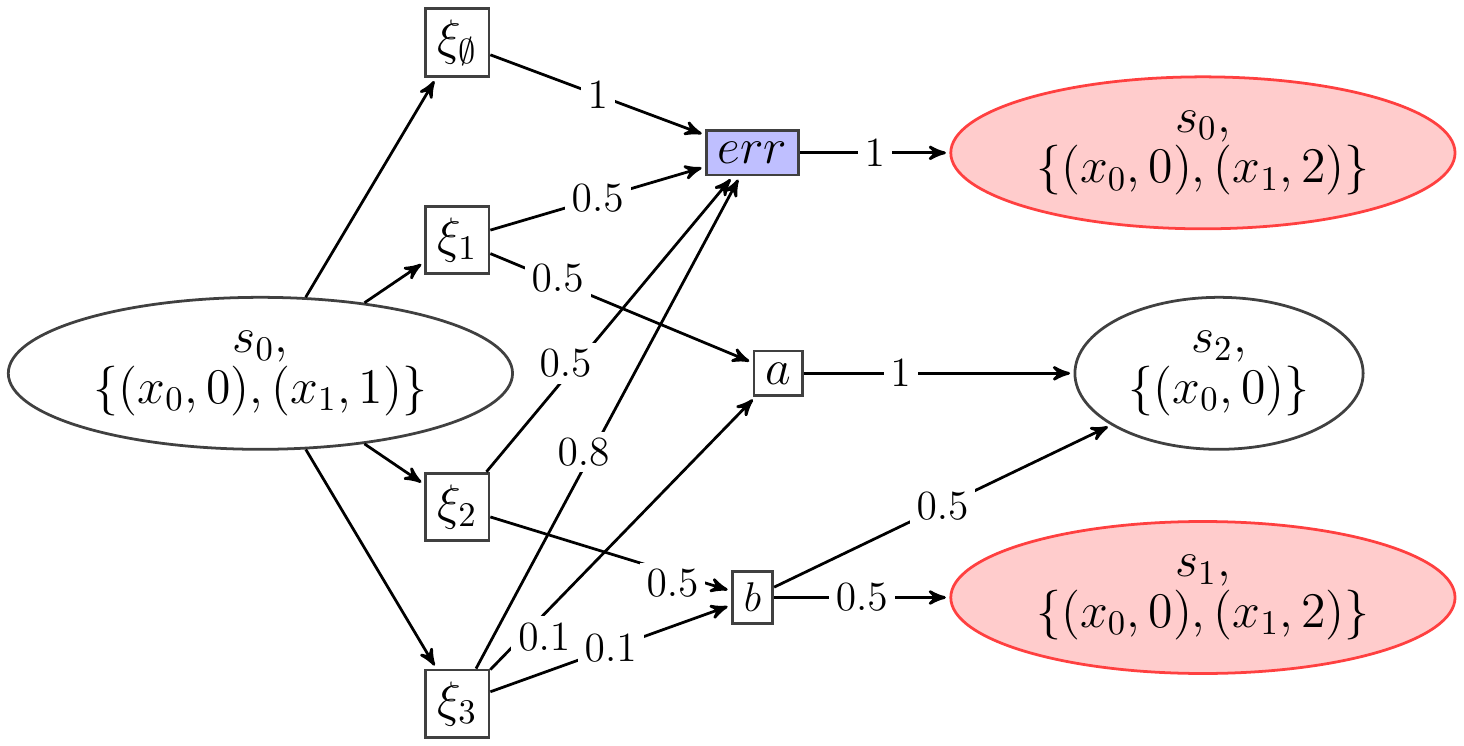}
  	\caption{A part of the product SDES from the SDES and the d$K$cBA with $K=1$ shown in Figs.\ \ref{SDES_S3} and \ref{fig:cBA_GFa}(b). Each accepting state is colored by red.}
  	\label{prodSDES}
  \end{figure}

For example, in Fig.\ \ref{prodSDES}, we show a part of the product SDES constructed from the SDES and the d$K$cBA with $K=1$ depicted in Figs.\ \ref{SDES_S3} and \ref{fig:cBA_GFa}(b).

\section{Supervised Stochastic Discrete Event Systems}
\label{SDES}
We consider a supervisory control problem of an SDES $D$ with an LTL formula $\varphi$. We synthesize a controller, called a \textit{supervisor}, to restrict the behavior to satisfy the specification. For $D$ and $\varphi$, without loss of generality, the supervisor is given by the pair of the determinized logical part of the corresponding product SDES\footnote{Since the transitions by the occurrence of the event in SDES is nondeterministic, we need the state after the transition to construct the supervisor.} and a mapping from its state to a control pattern \cite{RW1987}. Formally, for $D$ and $\varphi$, we define the supervisor as $\mathsf{SV}=(logic(D^\otimes),SV)$ where and $SV : Q \to \Xi$ is the mapping from a state to a control pattern.


We define the behavior of the SDES controlled by a supervisor.
Let $\mathsf{SV} = (logic(D^\otimes),$ $SV)$ be the supervisor for the SDES $D = (S, E, P_T, P_E, s^I, AP, L) $ under $\varphi$. 
The SDES $D$ controlled by $\mathsf{SV}$ is defined as $\mathsf{SV}/D = (\hat{S}, E, \hat{P}_T, \hat{P}_E, \hat s^I, AP, \hat{L})$, where $\hat S = S \times S^\otimes$, $\hat{P}_T : \hat{S} \times \hat{S} \times E \to [0,1]$ is defined as, for any $((s',s^{\otimes \prime}=(\tilde{s}, \tilde{F})), (s,s^\otimes), e) \in \hat{S} \times \hat{S} \times E$,
\begin{align}
  &\hat{P}_T( (s', s^{\otimes \prime}) | (s, s^\otimes), e) = \nonumber \\
  &\left\{
  \begin{aligned}
    &P_T(s^{\prime} | s, e) &   &\text{if}\  \tilde{s}=s', s^{\otimes \prime} = \delta(s^\otimes, (e,s^{\otimes \prime})), \\
    &0 &   &\text{otherwise} ,
  \end{aligned}
  \right. 
\end{align}
$\hat{P}_E : E \times \hat{S} \to [0,1]$ is defined as, for any $(e, (s,s^\otimes)) \in E \times \hat{S}$, $\hat{P}_E(e | (s,s^\otimes)) = P_E(e | s, SV(s^\otimes))$, $\hat s^I = (s^I, s^{\otimes I})$ is the initial state, $\hat{L} : \hat{S} \to 2^{AP}$ is defined as, for any $(s,s^\otimes) \in \hat{S}$, $\hat{L}((s,s^\otimes)) = L(s)$.

We consider a synthesis problem of a supervisor $\mathsf{SV}$ such that the controlled SDES $\mathsf{SV} / D$ satisfies $\varphi_B$. For convenience, we call the mapping $SV$ of $\mathsf{SV} = (logic(D^\otimes),SV)$ (resp., $SV^d$ of $\mathsf{SV}^d =(logic(D^\otimes),SV^d)$) the supervisor (resp., the directed controller).
In the following, for the SDES $D$ with the LTL specification $\varphi$, we focus on synthesizing a supervisor $SV : S^\otimes \to \Xi$ with $C=logic(D^\otimes)$.

We denote by $InfPath^D_{SV}(s)$ (resp., $FinPath^D_{SV}(s)$) the set of infinite (resp., finite) paths starting from $s \in S$ on $D$ under $SV$ and, for simplicity, $InfPath^D_{SV}(s^I)$ (resp., $FinPath^D_{SV}(s^I)$) is denoted by $InfPath^D_{SV}$ (resp., $FinPath^D_{SV}$). 


We denote the probability space obtained from $InfPath^D_{SV}(s)$ by $(InfPath_{SV}^D(s), \mathcal{F}_{InfPath^D_{SV}(s)}, Pr^D_{SV})$.
For any supervisor $SV$ and any state $s \in S$, the probability measure of all paths starting from $s$ that satisfies an LTL formula $\varphi$ on the SDES $D$ under $SV$ is defined as follows.

\begin{align}
Pr^{D}_{SV}(s \models \varphi) \!:=\! Pr^{D}_{SV}(\{ \rho \! \in \! InfPath^{D}_{SV}(s) \;|\; \rho \! \models \varphi\}).
\end{align}
Similarly, we define the probability of paths starting from the initial state $s$ and the initial directed control pattern $\xi_e$ satisfies $\varphi$ on $D$ under $SV$ as 
$Pr^{D}_{SV}(s,\xi_e \models \varphi).$
We call $Pr^{D}_{SV}(s^I \models \varphi)$ \textit{the satisfaction probability} of $\varphi$ on the SDES $D$ under $SV$. Intuitively, the satisfaction probability of $\varphi$ means that the probability of satisfying $\varphi$ from the initial state $s^I$ by $SV$ for $D$.

A Markov chain induced by the product SDES $D^\otimes$ with a supervisor $SV$ is a tuple $MC^\otimes_{SV} = (S^\otimes_{SV},P^\otimes_{SV},s^{\otimes I})$, where $S^\otimes_{SV} = S^\otimes$, $P^\otimes_{SV}(s'|s) = P(s'|s,SV(s))$ for $s, s^{\prime} \in S^\otimes$. The state set $S^\otimes_{SV}$ of $MC^\otimes_{SV}$ can be represented as a disjoint union of a set of transient states $T^\otimes_{SV}$ and closed irreducible sets of recurrent states $R^{\otimes j}_{SV}$ with $j \in \{ 1, \ldots ,h \}$, i.e., $ S^\otimes_{SV} = T^\otimes_{SV} \cup R^{\otimes 1}_{SV} \cup \ldots \cup R^{\otimes h}_{SV} $ \cite{MC_DMPS2018}. In the following, we say a ``recurrent class'' instead of a ``closed irreducible set of recurrent states'' for simplicity.

\begin{definition}[Reachable set]
For the set of states $S$ of an SDES $D$, the reachable set from a state $s \in S$ is defined as follows.
\begin{align}
    Re(S;s) = \{ last(\rho) \in S \;|\; \rho \in FinPaths^{D}(s) \}.
\end{align}
Moreover, we define the reachable set from $s \in S$ under a supervisor $SV$ as follows.
\begin{align}
    Re_{SV} (S;s) = \{ last(\rho) \in S \;|\; \rho \in FinPaths^{D}_{SV}(s) \}.
\end{align}
For simplicity, we denote the reachable set from the initial state $s^I$ by $Re(S)$ instead of $Re(S;s^{I})$. Analogously, for a supervisor $SV$, we write $Re_{SV}(S)$ instead of $Re_{SV}(S; s^{I})$. The reachable sets for a controlled SDES and a product SDES are defined in the same way.
\end{definition}

\begin{remark}
 For any controlled SDES $\mathsf{SV}/D$ and any reachable state $\hat s = (s, s^\otimes = (\tilde s, \tilde F)) \in Re( \hat S )$, $s=\tilde s$ holds.
\end{remark}

For the product SDES $D^\otimes$, let 
\begin{align}
    \label{W}
    W = \{ s \in S^\otimes \;|\; \exists SV \mbox{ s.t. } Pr^{D^\otimes}_{SV}(s \models \varphi_B) = 1\}.
\end{align}
$W$ is called the \textit{winning region} of $D^\otimes$.
Then, a supervisor $SV$ such that $Pr^{D^\otimes}_{SV}(s \models \varphi_B) =1$ is called a \textit{winning supervisor} at $s$. Moreover, it is called a \textit{winning directed controller} if it is a directed controller. 
Furthermore, let 
\begin{align}
    W_p = \{ (s,\xi) \!\in\! S^\otimes \times \Xi^\otimes_{dir} \;|\; \exists SV \mbox{ s.t. }  Pr^{D^\otimes}_{SV}(s,\xi \models \varphi_B) \!=\! 1 \}.
    \label{Wp}
\end{align}
$W_p$ is called the \textit{winning pair set} of $D^\otimes$. 

We define the inclusion relations between supervisors.
\begin{definition}[Inclusion relation]
Let $D^\otimes$ be a product SDES. For any supervisor $SV$ and $SV^\prime$, we say that $SV$ includes $SV^\prime$ if $SV^\prime(s) \subseteq SV(s)$ for any $s \in S^\otimes$ and denote $SV^\prime \subseteq_{D^\otimes} SV$ if $SV$ includes $SV^\prime$.
\end{definition}

We omit the subscript $D^\otimes$ of $\subseteq_{D^\otimes}$ for simplicity.

We define the notion of sure satisfaction.
\begin{definition}[Sure satisfaction of safety]
For a product SDES $D^\otimes$ and a supervisor $SV$, we say that $SV$ forces $D^\otimes$ to be safe at $s \in S^{\otimes}$ surely if
\begin{align}
\{ \rho \in InfPath_{SV}^{D^\otimes}(s) \;|\; D^\otimes, \rho \not \models \varphi_B \} = \emptyset.
\end{align}
\end{definition}
We denote by $\mathcal{SV}^{\otimes}_{sure}(s)$ the set of supervisors that force $D^\otimes$ to be safe surely at $s \in S^\otimes$.
For simplicity, we denote $\mathcal{SV}^{\otimes}_{sure}(s^I)$ by $\mathcal{SV}^{\otimes}_{sure}$. It is said that a supervisor $SV$ satisfies the acceptance condition of the product SDES $D^\otimes$ surely if $SV \in \mathcal{SV}_{sure}^{\otimes}$.

The following lemma shows an important property of supervisors for the product SDES $D^\otimes$.
\begin{lemma}
\label{lem_thm1}
Given a product SDES $D^\otimes$ of an SDES $D$ and a d$K$cBA $det(B,K)$ converted from a given LTL formula $\varphi$, the following three conditions are equivalent for any state $s \in S^\otimes$ and any supervisor $SV$.
\begin{enumerate}
    \item $SV \in \mathcal{SV}_{sure}^\otimes(s)$.
    \item $Pr_{SV}^{D^\otimes}(s \models \varphi_B) = 1$.
    \item $Pr_{SV^d}^{D^\otimes}(s \models \varphi_B) = 1$ for any directed controller $SV^d \subseteq SV$.
\end{enumerate}
\end{lemma}
\begin{IEEEproof} We fix a state $s \in S^\otimes$ arbitrarily.
\begin{LaTeXdescription}
    \item[\textnormal{1) $\Rightarrow$ 2)}] There is no path $\rho \in InfPath_{SV}^{D^\otimes}(s)$ such that $\rho \not\models \varphi_B$. Thus, we have $Pr_{SV}^{D^\otimes}(s \models \varphi_B) =1$.
    \item[\textnormal{2) $\Rightarrow$ 3)}] Suppose that there exists a directed controller $SV^d \subseteq SV$ such that $Pr_{SV^d}^{D^\otimes}(s \not\models \varphi) > 0$. Then, there exists $\rho =s_0 e_0 s_1 \ldots \in InfPath_{SV^d}^{D^\otimes}(s)$ such that $\rho \not\models \varphi_B$ and hence there exists $l \in \mathbb{N}$ such that $s_{l-1} \not\in Acc^\otimes$ and $s_l \in Acc^\otimes$. Note that $InfPath_{SV^d}^\otimes(s) \subseteq InfPath_{SV}^\otimes(s)$. Thus, we have $\rho \in InfPath_{SV}^{D^\otimes}(s)$ and hence $Pr_{SV}^{D^\otimes}(s \not\models \varphi_B) \geq \prod_{i=0}^{l-1} P^\otimes_T(s_{i+1}|s_i,e_i)P_E(e_i|s_i,SV(s_i)) > 0$. This contradicts $Pr_{SV}^{D^\otimes}(s \models \varphi_B) = 1$.
    \item[\textnormal{3) $\Rightarrow$ 1)}] Suppose that $SV \not\in \mathcal{SV}^\otimes_{sure}(s)$. Then, there exists a path $\rho = s_0 e_0 s_1 \ldots \in InfPath_{SV}^\otimes(s)$ such that $\rho \not \models \varphi_B$. Then without loss of generality, we can assume that  $\rho=s_0e_0s_1 \ldots s_{l-1}e_{l-1}s_{l} e_{l}s_{l+1} \ldots$ such that $s_{l} \in Acc^{\otimes}$ and, for any $i$, $j$ with $0\leq i < j \leq l$, $s_i \not= s_j$.   Hence, there exists $SV^d \subseteq SV$ such that $\rho \in InfPath_{SV^d}^{\otimes}(s)$. Thus, we have $Pr^\otimes_{SV^d}(s \not\models \varphi_B) \geq \prod_{i=0}^{l-1}P^\otimes_T(s_{i+1}|s_i,e_i)P^\otimes_E(e_i|s_i,SV^d(s_i)) >0 $. However, this contradicts the assumption that $Pr_{SV^d}^{D^\otimes}(s \models \varphi_B) = 1$ for any $SV^d \subseteq SV$.
\end{LaTeXdescription}
\end{IEEEproof}



\section{Bounded Synthesis of Directed controllers}
\label{BSS}

In this section, for a given SDES $D$, a given LTL formula $\varphi$, and a non-negative integer $K$, we aim at the synthesis of directed controllers $SV^d$ for the product SDES $D^{\otimes}=D \otimes det(B_{\varphi}, K)$ such that $Pr_{SV^d}^{D^{\otimes}}(s^{\otimes I} \models \varphi_{B})$ is maximized, where $s^{\otimes I}$ and $Acc^{\otimes}$ are the initial state and the accepting set of $D^{\otimes}$.
We propose a synthesis method based on the bounded synthesis \cite{FJR2011, SF2007}. In the following, we first define a reward function and an expected return. Next, we show some important properties between the expected return and directed controllers for the product SDES. Finally, we give a dynamic programming based method to obtain an optimal directed controller.

\subsection{Reward function and Expected Returns}
\label{PSaER}
  We define a reward function based on the $K$-co-B\"uchi acceptance condition of the product SDES.


\begin{definition}[Reward function]\label{reward_def}
  The reward function $\mathcal{R} : S^\otimes \to \mathbb{R}$ is defined by

    \begin{align}
    \mathcal{R} (s) =
    \left\{
    \begin{aligned}
      & (1 - \gamma_{acc})r_n & &\text{if} \ s \in Acc^\otimes , \\
      & 0 & &\text{otherwise},
    \end{aligned}
    \right.
    \label{def_reward}
  \end{align}
  where $r_n$ is a negative value and $\gamma_{acc} \in (0,1)$.
\end{definition}

We introduce a discount factor function and value functions for the product SDES inspired by \cite{BWZP2020}.

\begin{definition}[Discount factor function]
\label{discount_def}
  A discount factor function $\Gamma : S^{\otimes} \rightarrow \{ \gamma_{acc}, \gamma \} $, where $\gamma \in (0,1)$, is defined by

  \begin{align}
      \Gamma (s) =
    \left\{
    \begin{aligned}
      & \gamma_{acc} & & \text{if}\ s\in Acc^{\otimes} \!,\\
      & \gamma & & \text{otherwise}.
    \end{aligned}
    \right.
    \label{def_discount}
  \end{align}
\end{definition}

\begin{definition}[Expected return]
\label{Exp_returns}
  For a directed controller $SV^d$ for the product SDES $D^\otimes$, the reward function $\mathcal{R}$ defined as (\ref{def_reward}), and the discount factor function $\Gamma$ defined as (\ref{def_discount}), we define the expected return, or the state-value function $V_{SV^d} : S^\otimes \to \mathbb{R}$ as follows. For each $s \in S^\otimes$,
  \begin{align}
   \label{V_def}
    V_{SV^d}(s)= \mathbb{E}_{SV^d}[\sum_{t=0}^{\infty}\mathcal{R}(s_{t+1}) \prod_{k=0}^{t-1} \Gamma(s_{k+1})|s_0 = s],
  \end{align}
where $\mathbb{E}_{SV^d}$ denotes the expected value of accumulated rewards from the state $s$ under the directed controller $SV^d$ and $\prod_{k=0}^{-1} \Gamma(s_{k+1}) := 1$.
\end{definition}

Note that an upper and a lower bound of the state-value function $V_{SV^d}$ are $0$ and $r_{n}$, respectively, and $V_{SV^d}(s)=r_n$ for any unsafe state $s \in Acc^{\otimes}$. 
We define a function $Q_{SV^d} : \{ (s,\xi_e) | s \in S^\otimes, \xi_e \in \Xi^\otimes_{dir}(s) \} \to \mathbb{R}$ as follows. For any $s \in S^\otimes$ and any $\xi_e \in \Xi^\otimes_{dir}(s)$, 
  \begin{align}
    \lefteqn{Q_{SV^d}(s, \xi_e)}\nonumber \\
    &=\mathbb{E}_{SV^d}[\sum_{t=0}^{\infty} \mathcal{R}(s_{t+1})\prod_{k=0}^{t-1} \Gamma(s_{k+1})|s_0 = s, \xi_0 = \xi_e],
    \label{Q_def}
  \end{align}
where $\mathcal{R}$ and $\Gamma$ are defined as (\ref{def_reward}) and (\ref{def_discount}), respectively. We call $Q_{SV^d}$ a state directed control pattern (state-DCP for short) value function under $SV^d$.

From the state and state-DCP value functions, we have the following \textit{Bellman equations} \cite{SB2018}.

  \begin{align*}
    &V_{SV^d}(s) =\sum_{s^{\prime} \in S^\otimes}P^\otimes(s^{ \prime}|s, SV^d(s)) \{ \mathcal{R}(s^{ \prime}) + \Gamma(s) V_{SV^d}(s^{ \prime}) \}, \\
    &Q_{SV^d}(s, \xi_e) \\
    & = \sum_{s^{\prime} \in S^\otimes}P^\otimes(s^{ \prime}|s, \xi_e) \{ \mathcal{R}(s^{ \prime}) + \Gamma(s) Q_{SV^d}(s^{ \prime},SV^d(s^\prime)) \}. 
  \end{align*}

\begin{definition}[Optimal value functions and optimal directed controllers]
  Let $D^\otimes$ be a product SDES. The state value function $V^\ast$ is the optimal state value function if, for any $s \in S^\otimes$, 
  \begin{align}
   V^\ast(s) = \max_{SV^d} V_{SV^d}(s).
  \end{align}
  Likewise, the state-DCP value function $Q^\ast$ is optimal if, for any $s \in S^\otimes$ and $\xi_e \in \Xi^\otimes_{dir}(s)$,
  \begin{align}
      Q^{\ast}(s,\xi_e) = \max_{SV^d} Q_{SV^d}(s,\xi_e)
  \end{align}
  Then, a directed controller $SV^{d\ast}$ is called an optimal directed controller if, for any $s \in S^\otimes$,
  \begin{align}
      \label{SVdast}
      V_{SV^{d\ast}}(s) = V^\ast(s).
  \end{align}
\end{definition}  
  
It is shown that the following \textit{optimal Bellman equations} hold \cite{SB2018}.
\begin{align}
    Q^{\ast}(s,\xi_e) &= \sum_{s^{\prime} \in S^\otimes}P^\otimes(s^{ \prime}|s, \xi_e) \{ \mathcal{R}(s^{ \prime}) \nonumber \\
      &  \hspace{20mm}  + \Gamma(s) \max_{\xi_{e^\prime} \in \Xi_{dir}^\otimes(s^\prime)} Q^\ast(s^{ \prime},\xi_{e^\prime}) \}, \label{Q_opt_Bellman} \\
     V^{\ast}(s) &= \max_{\xi_{e} \in \Xi_{dir}^\otimes(s)} Q^{\ast}(s,\xi_e) \nonumber \\
                 &= \max_{\xi_{e} \in \Xi_{dir}^\otimes(s)}\! \sum_{s^{\prime} \in S^\otimes}P^\otimes(s^{ \prime}|s, \xi_e) \{ \mathcal{R}(s^{ \prime}) \! +\! \Gamma(s) V^\ast(s^{ \prime}) \}. 
\end{align}

\subsection{Properties between Directed controllers and Expected return}
\label{BSod}

For any product SDES $D^\otimes$ associated with an LTL formula $\varphi$ and any supervisor $SV$, let $S^\otimes_{SV \not\models} = \{ s \in S^\otimes | \exists \rho \in InfPath^{D^\otimes}_{SV}(s) \mbox{ s.t. } \rho \not\models \varphi_B\}$. Note that $S^\otimes_{SV \not\models}$ is non-empty since it contains all unsafe states. Moreover, let $N_\rho = \min \{ n \in \mathbb{N}_0 | \rho[n]\in Acc^\otimes \} $ for each $\rho \in InfPath^{D^\otimes}_{SV}(s)$. We define $\min \emptyset = \infty$. Then, we show the following two lemmas.
\begin{lemma}
Given a product SDES $D^\otimes$ associated with an LTL formula $\varphi$, for any supervisor $SV$, $\max_{s \in S^\otimes_{SV \not\models}} \mathbb{E}_{SV} [N_\rho | \rho[0] = s, \rho \not\models \varphi_B] < \infty$.
\label{lem_N_fin}
\end{lemma}
\begin{IEEEproof}
  For any supervisor $SV$, any state $s \in S^\otimes_{SV \not\models}$, and any path $\rho \in InfPath^{D^\otimes}_{SV}(s)$ with $\rho \models \textbf{F}Acc^\otimes$, we have $N_\rho < \infty$ since there exists a successor state of $\rho[N_\rho]$ in $Acc^\otimes$. Thus, we have $\mathbb{E}_{SV}[N_\rho | \rho[0] =s, \rho \not\models \varphi_B ] < \infty$. Therefore, we have $\max_{s\in S^\otimes_{SV \not\models}} \mathbb{E}_{SV}[N_\rho | \rho[0] =s, \rho \not\models \varphi_B ] < \infty$.
\end{IEEEproof}

\begin{lemma}
\label{lemma_connect_V_Pr}
Given a product SDES $D^\otimes$ associated with an LTL formula $\varphi$, a directed controller $SV^d$, and the reward function $\mathcal{R}$ defined as (\ref{def_reward}), the following two conditions hold for any discount factor $\gamma \in (0,1)$.
\begin{enumerate}
    \item For any $s \in S^\otimes$, $V_{SV^d}(s) = 0$ if and only if $Pr_{SV^d}^{D^\otimes}(s \models \varphi_B) = 1$.
    \item There exists a constant $N$ such that, for any $s \in S^\otimes$, 
    \begin{align}
        r_nPr^{D^\otimes}_{SV^{d}}(s \not\models \varphi_B) \leq V_{SV^d}(s) < r_nPr^{D^\otimes}_{SV^{d}}(s \not\models \varphi_B) \gamma^{N}
        \label{V_inequality}
    \end{align}
    if and only if $Pr_{SV^d}^{D^\otimes}(s \models \varphi_B) < 1$.
\end{enumerate}
\end{lemma}

\begin{IEEEproof}
For any $s \in S^\otimes$ and any $\gamma \in (0,1)$, we have
 \begin{align}
    V_{SV^{d}}(s) \nonumber
    &= \mathbb{E}_{SV^{d}} [ \sum_{t=0}^{\infty} \mathcal{R}(s_{t+1}) \prod_{k=0}^{t-1} \Gamma(s_{k+1}) | s_0 = s ]\\ \nonumber
    &= \mathbb{E}_{SV^{d}} [ \sum_{t=0}^{\infty} \mathcal{R}(s_{t+1}) \prod_{k=0}^{t-1} \Gamma(s_{k+1}) | \rho[0] = s, \rho \not\models \varphi_B ]\\ \nonumber
    & \hspace{5mm} \times Pr^{D^\otimes}_{SV^{d}}(s \not\models \varphi_B)\\ \nonumber
    &= \mathbb{E}_{SV^{d}} [ \gamma^{N_{\rho}-1} \sum_{t=0}^{\infty}  \gamma_{acc}^t \mathcal{R}(s_{N_{\rho}+t}) | \rho[0] = s, \rho \not\models \varphi_B ]\\ \nonumber
    & \hspace{5mm} \times Pr_{SV^{d}}(s \not\models \varphi_B). \nonumber
    \intertext{Thus, we have}
    V_{SV^{d}}(s) &= r_nPr^{D^\otimes}_{SV^{d}}(s \not\models \varphi) \mathbb{E}_{SV^{d}}[ \gamma^{N_{\rho}-1}| \rho[0] = s, \rho \not\models \varphi_B ].
    \label{V_Pr}
\end{align}
We now prove 1).
\begin{LaTeXdescription}
  \item[\textnormal{($\Leftarrow$)}] By (\ref{V_Pr}), we have $V_{SV^d}(s) = 0$.
  \item[\textnormal{($\Rightarrow$)}] Suppose $Pr^{D^\otimes}_{SV^d}(s \not\models \varphi_B) > 0$. Then, by Lemma \ref{lem_N_fin}, $\mathbb{E}[N_\rho - 1 | \rho[0]=s, \rho \not\models \varphi_B] < \infty$. Recall that the reward $r_n<0$. By Jensen's inequality, $V_{SV^d}(s) \leq r_n Pr^{D^\otimes}_{SV^d}(s \models \varphi)\gamma^{\mathbb{E}[N_\rho - 1 | \rho[0]=s, \rho \not\models \varphi_B ]} < 0$. This contradicts $V_{SV^d}(s)=0$.
\end{LaTeXdescription}

Next, we prove 2).
\begin{LaTeXdescription}
  \item[\textnormal{($\Leftarrow$)}] Let $N = \max_{s \in S^\otimes_{SV^d \not\models}} \mathbb{E}[N_\rho | \rho[0]=s, \rho \not\models \varphi_B]$. Recall that $r_n<0$. Note that $N < \infty$ by Lemma \ref{lem_N_fin} and it is independent of the states. Moreover, $\gamma^{\mathbb{E}[N_\rho - 1 | \rho[0]=s, \rho \not\models \varphi_B ]} > \gamma^N > 0$ since $\gamma \in (0,1)$. By Jensen's inequality, (\ref{V_inequality}) holds.
  \item[\textnormal{($\Rightarrow$)}] Suppose $Pr^{D^\otimes}_{SV^d}(s \not\models \varphi_B) = 0$. Then, we have $V_{SV^d}(s) = Pr^{D^\otimes}_{SV^d}(s \not\models \varphi_B) = 0$ by (\ref{V_Pr}). This contradicts the existence of the constant $N$ that satisfies (\ref{V_inequality}).
\end{LaTeXdescription}
\end{IEEEproof}

We now show that an optimal directed controller is winning at any state in the winning region $W$. 

\begin{theorem}
\label{thm_SVdast_winning}
Given a product SDES $D^\otimes$ associated with an LTL formula $\varphi$ and the reward function $\mathcal{R}$ defined as (\ref{def_reward}), for the winning region $W$ of $D^\otimes$ and any optimal direted controller $SV^{d\ast}$, we have \begin{align}
 \label{SVast_winning}
    Pr^{D^\otimes}_{SV^{d\ast}}(s \models \varphi_B) = 1 \mbox{ for any } s \in W
\end{align}
\end{theorem}

\begin{IEEEproof}
 By (\ref{W}), 2) and 3) of Lemma \ref{lem_thm1}, and 1) of Lemma \ref{lemma_connect_V_Pr}, for any $s \in W$, there exists a directed controller $SV^d$ such that $V_{SV^d}(s) = 0$. Thus, by (\ref{SVdast}), we have $V^\ast(s) = 0$ for any $s \in W$. Hence, by 1) of Lemma \ref{lemma_connect_V_Pr}, for any optimal directed controller $SV^{d\ast}$, (\ref{SVast_winning}) holds.
\end{IEEEproof}

The following theorem shows the existence of a directed control pattern for which the state-DCP value is 0 at a state is closely related to the satisfaction of the given LTL formula from the state.
\begin{theorem}
\label{th_equivalence}
Given a product SDES $D^\otimes$ associated with an LTL formula $\varphi$ and its state $s \in S^\otimes$, there exists a supervisor $SV$ such that $Pr_{SV}^{D^\otimes}(s \models \varphi_B) = 1$ if and only if there exists $\xi_e \in \Xi_{dir}^\otimes(s)$ such that $Q^\ast(s,\xi_e) = 0$.
\end{theorem}

\begin{IEEEproof}
 \begin{LaTeXdescription}
    \item[\textnormal{($\Leftarrow$)}] By 1) of Lemma \ref{lemma_connect_V_Pr}, we have $Pr_{SV^{d\ast}}^{D^\otimes}(s \models \varphi_B) = 1$, where $SV^{d\ast}$ is an optimal directed controller with $SV^{d\ast}(s) = \xi_e$.
    \item[\textnormal{($\Rightarrow$)}]
    By 2) and 3) of Lemma \ref{lem_thm1}, for any directed controller $SV^d \subseteq SV$, we have $Pr_{SV^d}^{D^\otimes}(s \models \varphi_B) = 1$. Thus, by 1) of Lemma \ref{lemma_connect_V_Pr}, we have $Q_{SV^d}(s,SV^d(s)) = 0$, which implies that $Q^\ast(s,SV^d(s))=0$.
 \end{LaTeXdescription}
\end{IEEEproof}

\begin{remark}
Theorem \ref{th_equivalence} implies that, if $Q^\ast(s^I, \xi_e) < 0$ for any $\xi_e \in \Xi_{dir}^\otimes(s^I)$, then there is no supervisor that satisfies the acceptance condition of the product SDES with probability 1. In general, supervisors constructed via the value function based on the directed control patterns can fail to accomplish the maximum satisfaction probability in such cases, for instance, see Example 1 in \cite{JJDTK2016}.
Furthermore, the occurrence probabilities of events can change depending on given control patterns.
Taking these issues into account, we will construct a supervisor in Section \ref{MPSC}.
\end{remark}

In the following lemma, we show that, for any state of a given product SDES,  its state value converges to the negative reward $r_n$ multiplied by the probability of reaching an unsafe state from the state as the discount factor $\gamma$ goes to 1. The lemma follows from similar proof as Theorem 1 in \cite{BWZP2020}.

\begin{lemma}
Given a product SDES $D^\otimes$ associated with an LTL formula $\varphi$, and the reward function $\mathcal{R}$ defined as (\ref{def_reward}), for any $\varepsilon > 0$, there exists $\gamma' > 0$ such that, for any $\gamma > \gamma'$, any state $s \in S^\otimes$, and any directed controller $SV^d$, the following equation holds.
\begin{align}
\label{V_Pr_ineq}
    0 \leq V_{SV^d}(s) - r_n Pr_{SV^d}(s \not\models \varphi_B) < \varepsilon.
\end{align}
\label{lemma_directed_main}
\end{lemma}

\begin{IEEEproof}
 For any state $s \in S^\otimes$ and any directed controller $SV^d$, consider the following two cases 1) $Pr_{SV^d}^{D^\otimes}(s \not\models \varphi_B) = 0$ and 2) $Pr_{SV^d}^{D^\otimes}(s \not\models \varphi_B) > 0$. For the case 1), we have $V_{SV^d}(s) = 0$ by 1) of Lemma \ref{lemma_connect_V_Pr}. For the case 2), by 2) of Lemma \ref{lemma_connect_V_Pr}, we have
\begin{align*}
    0 &\leq V_{SV^d}(s) - r_n Pr^{D^\otimes}_{SV^d}(s \not\models \varphi) \\
    &\leq r_nPr^{D^\otimes}_{SV^{d}}(s \not\models \varphi) \gamma^N - r_nPr^{D^\otimes}_{SV^{d}}(s \not\models \varphi) \\
    &= |r_n| Pr^{D^\otimes}_{SV^{d}}(s \not\models \varphi) ( 1 - \gamma^{N} )\\
    &\leq |r_n| ( 1 - \gamma^{N} ),
\end{align*}
where $N$ is a constant that satisfies 2) of Lemma \ref{lemma_connect_V_Pr}. Thus, for any $\varepsilon > 0$, there exists $\gamma' \in (0,1)$ such that, for any $\gamma > \gamma'$, $|r_n| ( 1 - \gamma^{N} ) < \varepsilon$. Note that $\gamma'$ is independent of $s$ and $SV^d$. Therefore, (\ref{V_Pr_ineq}) holds for any $\gamma > \gamma'$, any state $s \in S^\otimes$, and any directed controller $SV^d$.
\end{IEEEproof}

For any product SDES $D^\otimes$ associated with an LTL formula $\varphi$, any $s \in S^\otimes$, and any directed controller $SV^d$, we define
\begin{align}
    Pr^{D^\otimes}_{max^d}(s \models \varphi_B) = \max_{SV^d} Pr^{D^\otimes}_{SV^d}(s \models \varphi_B).
    \label{def_max_prob}
\end{align}
Similarly, we define
\begin{align}
    Pr^{D^\otimes}_{max^d}(s, \xi_e \models \varphi_B) = \max_{SV^d} Pr^{D^\otimes}_{SV^d}(s, \xi_e \models \varphi_B).
    \label{def_max_prob_xie}
\end{align}
By Lemma \ref{lemma_directed_main}, we now show that the satisfaction probability of an optimal directed controller is exactly equal to the maximum satisfaction probability when the discount factor $\gamma$ is sufficiently close to 1. This plays an important role to assure that the supervisor constructed in Section \ref{MPSC} achieves the maximum satisfaction probability.

\begin{theorem}
\label{thm_directed_main}
Given a product SDES associated with an LTL formula $\varphi$ and the reward function $\mathcal{R}$ defined as (\ref{def_reward}), there exists a discount factor $\gamma' \in (0,1)$ such that, for any $\gamma > \gamma'$ and any $s \in S^\otimes$, the following equation holds. 
\begin{align}
 \label{Prast_Prmax_equi}
    Pr^{D^\otimes}_{SV^{d\ast}}(s \models \varphi_B) = Pr^{D^\otimes}_{max^d}(s \models \varphi_B).
\end{align}
\end{theorem}

\begin{IEEEproof}
 Note that the number of directed controllers is finite since $|E_c| < \infty$ and $|S^\otimes| < \infty$. Thus, there exists $\varepsilon > 0$ such that, for any non-optimal directed controller $SV^d$, there exists a state $s \in S^\otimes$, we have
\begin{align}
  V_{SV^d}(s) + \varepsilon < V^\ast(s).
  \label{ineq1_in_thm_Prast_Prmax}
 \end{align}
 Then, by Lemma \ref{lemma_directed_main}, there exists $\gamma' \in (0,1)$ such that, for any $\gamma > \gamma'$,
 \begin{align}
 V^\ast(s) - r_n Pr^{D^\otimes}_{SV^{d\ast}}(s \not\models \varphi_B) < \varepsilon.
  \label{ineq2_in_thm_Prast_Prmax}
 \end{align}
 By (\ref{ineq1_in_thm_Prast_Prmax}) and (\ref{ineq2_in_thm_Prast_Prmax}), we have 
 \begin{align*}
   V_{SV^d}(s) &< r_n Pr^{D^\otimes}_{SV^{d\ast}}(s \not\models \varphi_B).
 \end{align*}
 Note that $r_n$ is negative. Hence, by (\ref{V_inequality}), we have
 \begin{align*}
   Pr^{D^\otimes}_{SV^{d}}(s \not\models \varphi_B) > Pr^{D^\otimes}_{SV^{d\ast}}(s \not\models \varphi_B).
 \end{align*}
 Therefore, we have 
 \begin{align}
 Pr^{D^\otimes}_{SV^{d}}(s \not\models \varphi_B) > Pr^{D^\otimes}_{SV^{d\ast}}(s \not\models \varphi_B)
 \quad \mbox{for any } SV^d,
 \end{align}
 which implies together with (\ref{def_max_prob}) that (\ref{Prast_Prmax_equi}) holds.
\end{IEEEproof}

By Theorem \ref{thm_directed_main}, we obtain a directed controller with the maximum satisfaction probability by maximizing the state value function when the discount factor $\gamma$ exceeds a certain value $\gamma'$ that is less than 1. 

When we know the full information about the dynamics of the SDES, to compute the directed controllers achieving the maximum satisfaction probability, we can use a dynamic programming (DP) method such as the value iteration. We show the value iteration with directed controllers in Algorithm \ref{vi}\footnote{We use the value iteration here but any DP method can be employed to our proposed framework.}. Note that $\Gamma(s) < 1$ for any state $s \in S^\otimes$ and thus the Bellman operator corresponding to (\ref{Q_opt_Bellman}) is a contraction mapping. This implies that, for any product SDES $D^\otimes$ and any state $s \in S^\otimes$, initializing the state-DCP value function $Q$ with 0, if there exists a winning directed controller $SV^d$, then $Q(s,SV^d(s))$ is always 0 during the value iteration. Moreover, Algorithm \ref{vi} converges exponentially.

\begin{algorithm}
   \caption{Value iteration with directed control.}
   \label{vi}
   \begin{algorithmic}[1]
   \renewcommand{\algorithmicrequire}{\textbf{Input:}}
   \renewcommand{\algorithmicensure}{\textbf{Output:}}
   \REQUIRE LTL formula $\varphi$ and SDES $D$
   \ENSURE An optimal supervisor $SV^\ast$.
   \STATE Convert $\varphi$ to d$K$cBA $\det(B_{\varphi},K)$.
    \STATE Construct the product SDES $D^{\otimes}$ of $D$ and $det(B_{\varphi},K)$.
    \STATE Initialize $Q:S^{\otimes} \times E^{\otimes} \to \mathbb{R}$ with 0.
    \WHILE {$Q$ does not converge}
    \FOR {all state $s \in S^\otimes$ and $\xi_e \in \Xi^\otimes_{dir}(s)$}
    \STATE \begin{varwidth}[t]{\linewidth}
    \begin{align*}
        Q(s^{\otimes},\xi_e) & \gets \sum_{s^{\prime} \in S^\otimes}P^\otimes_T(s^{ \prime}|s^\otimes, \hat{e}) \sum_{\hat{e} \in \xi_e} P^\otimes_E(\hat{e}|s, \xi_e)\\
        & \times \{ \mathcal{R}(s^{ \prime}) + \Gamma(s^{ \prime}) \max_{\xi_{e^\prime} \in \Xi^\otimes_{dir}(s^{ \prime})}Q(s^{ \prime},\xi_{e^\prime}) \}.
    \end{align*}
    \end{varwidth}
    \ENDFOR
    \ENDWHILE
   \end{algorithmic}
  \end{algorithm}

  \begin{example}
  \label{ex1}
  We consider the SDES depicted in Fig.\ \ref{SDES_S3} and the d$K$cBA shown in Fig.~\ref{fig:cBA_GFa}(2), i.e., the d$K$cBA with $K=1$ for $\varphi=\textbf{GF}r$. A part of the corresponding product SDES is shown in Fig.\ \ref{prodSDES}. The initial state is $s^{\otimes I} = (s_1, \{(x_0,0),(x_1,1)\})$. We set $\gamma = 0.9999$, $\gamma_{acc} = 0.9$, and $r_n=-1$. We apply Algorithm \ref{vi} on the product SDES. The obtained state-DCP values are shown in Table \ref{tabale_ex1} and only the states in the column are reachable. For convenience, we abbreviate $err$ from each directed control pattern in Table \ref{tabale_ex1} when the directed control pattern contains a controllable event. Only ($s_1$, \{($x_0$,0),($x_1$,1)\}) and ($s_2$, \{($x_0$,0)\}) have state-DCP values of 0. Thus, the winning region is $\{ (s_1, \{(x_0,0),(x_1,1)\}), (s_2, \{(x_0,0)\}) \} $. Moreover, the absolute values of the state DCP values are close to their probability of reaching an unsafe state. For example, at the state $(s_1, \{(x_0,0),(x_1,1)\})$, the probability of reaching an unsafe state under the directed control pattern $\xi_1 = \{ a, err \}$ and $\xi_2 = \{b, err \}$ are 0.5 and 0.75, respectively. Note that the states except for the two states are not reachable from the initial state under any optimal directed controller.

  \begin{table}[htbp]
  \caption{The state-DCP values for Example \ref{ex1} obtained by Algorithm \ref{vi}.}
  \label{tabale_ex1}
  \centering
    \begin{tabular}{ c||c c c c} \hline
       \diagbox{state}{DCP} & \{$a$\} & \{$b$\} & \{c\} & \{$err$\} \\ \hline \hline
      ($s_1$, \{($x_0$,0),($x_1$,1)\}) & -0.9999 & 0 & -- & -- \\ 
      ($s_1$, \{($x_0$,0),($x_1$,2)\}) & -1 & -1 & -- & -- \\ 
      ($s_2$, \{($x_0$,0)\}) & 0 & -0.4999 & 0 & -- \\ 
      ($s_0$, \{($x_0$,0),($x_1$,1)\}) & -0.49995 & -0.74992 & -- & -0.9999  \\ 
      ($s_0$, \{($x_0$,0),($x_1$,2)\}) & -1 & -1 & -- & -1\\ \hline
    \end{tabular}
  \end{table}
  \end{example}

We will show that, under any optimal directed controller, the satisfaction probability is equal to the probability of reaching the winning region. To establish that, we show the following lemma.

\begin{lemma}
 Given a product SDES $D^\otimes$ associated with an LTL formula $\varphi$ and the reward function $\mathcal{R}$ defined as (\ref{def_reward}), for any optimal directed controller $SV^{d\ast}$ and any state $s \in S^\otimes$, the following equations hold.
\begin{align}
&Pr^{D^\otimes}_{SV^{d\ast}}(s \models \varphi_B \land \textbf{F} W) = Pr^{D^\otimes}_{SV^{d\ast}}(s \models \textbf{F} W), \label{GnAcc_FWinf_FWinf} \\
&Pr^{D^\otimes}_{SV^{d\ast}}(s \models \varphi_B \land \textbf{G} \neg W) = 0,
\label{GnAcc_GnWinf_0}
\end{align}
where $W$ is the winning region of $D^\otimes$.

\end{lemma}

\begin{IEEEproof}
 We choose an optimal directed controller $SV^{d\ast}$ arbitrarily.
 To establish (\ref{GnAcc_FWinf_FWinf}), it is sufficient to show that, for any $s \in S^\otimes$ and any $\rho  \in InfPath^{D^\otimes}_{SV^{d\ast}}(s)$, $\rho \models \textbf{F}W$ implies $\rho \models \varphi_B \land \textbf{F}W$. Fix a state $s \in S^\otimes$ and a path $\rho \in InfPath^{D^\otimes}_{SV^{d\ast}}(s)$ arbitrarily. Suppose that $\rho \models \textbf{F}W$. Then, by Theorem \ref{thm_SVdast_winning} and 1) and 2) of Lemma \ref{lem_thm1}, we have $\rho \models \varphi_B$. Thus, for any optimal directed controller $SV^{d\ast}$ and any $s \in S^\otimes$, (\ref{GnAcc_FWinf_FWinf}) holds.
 
 Let $MC^\otimes_{SV^{d\ast}} = (S^\otimes, P^\otimes_{SV^{d\ast}}, s^{I\otimes}, AP, L)$ be the Markov chain induced by $D^\otimes$ and $SV^{d\ast}$. $S^{\otimes}$ is partitioned into a transient states $T^\otimes_{SV^{d\ast}}$ and $h$ recurrent classes $\{R^{i\otimes}_{SV^{d\ast}}\}_{i=1}^h$, that is $S^\otimes = T^\otimes_{SV^{d\ast}} \cup R^{1\otimes}_{SV^{d\ast}} \cup \ldots \cup R^{h\otimes}_{SV^{d\ast}}$.
 Suppose that there exists a state $s \not\in Acc^\otimes \cup W$. Then, we have $Pr^{D^\otimes}_{SV^{d\ast}}(s \models \varphi_B) < 1$ by the definition of $W$. Thus, there exists a path that eventually reaches $Acc^\otimes$ from $s$ on $MC^\otimes_{SV^{d\ast}}$. Recall that $Acc^\otimes$ has no outgoing transition to $S^\otimes \setminus Acc^\otimes$. Hence, we have that the probabilities of reaching $Acc^\otimes$ from $s$ and returning to $s$ from $Acc^\otimes$ are positive and 0, respectively, which implies $s \in T^\otimes_{SV^{d\ast}}$. Therefore, by the property of transient states \cite{MC_DMPS2018}, (\ref{GnAcc_GnWinf_0}) holds.
\end{IEEEproof}

\begin{prop}
\label{prop_phiB_FW_direct}
Given a product SDES $D^\otimes$ associated with an LTL formula $\varphi$ and the reward function $\mathcal{R}$ defined as (\ref{def_reward}), for any optimal directed controller $SV^{d\ast}$ and any state $s \in S^\otimes$, the following equation holds.
\begin{align}
\label{phiB_equalto_W}
Pr^{D^\otimes}_{SV^{d\ast}}(s \models \varphi_B) = Pr^{D^\otimes}_{SV^{d\ast}}(s \models \textbf{F} W).
\end{align}
\end{prop}

\begin{IEEEproof}
For any $s \in S^\otimes$ and any optimal directed controller $SV^{d\ast}$, we have
\begin{align*}
    Pr^\otimes_{SV^{d\ast}}(s \models \varphi_B) = Pr^\otimes_{SV^{d\ast}}(s \models \varphi_B \land (\textbf{F}W \lor \textbf{G}\neg W) ).
\end{align*}
By (\ref{GnAcc_FWinf_FWinf}) and (\ref{GnAcc_GnWinf_0}), we have
\begin{align*}
    &Pr^\otimes_{SV^{d\ast}}(s \models \varphi_B \land (\textbf{F}W \lor \textbf{G}\neg W) ) \\
    &= Pr^{D^\otimes}_{SV^{d\ast}}(s \models \textbf{F}W) + Pr^{D^\otimes}_{SV^{d\ast}}(s \models \varphi_B \land \textbf{G}\neg W)\\
    &= Pr^{D^\otimes}_{SV^{d\ast}}(s \models \textbf{F}W).
\end{align*}
Thus, (\ref{phiB_equalto_W}) holds.
\end{IEEEproof}

Proposition \ref{prop_phiB_FW_direct} together with Lemma \ref{lem_thm1} allows us to develop a supervisor that maximizes the number of reachable states within the winning region under the satisfaction of $\varphi_B$ with the maximum probability.
  
\section{Construction of Permissive Supervisor}
\label{MPSC}

 By the discussions in Sections \ref{SDES} and \ref{BSS}, an optimal directed controller is winning at any state in the winning region $W$. Thus, a supervisor that includes only optimal directed controllers surely satisfies $\varphi_B$ at any state in $W$. Besides, an optimal directed controller achieves the maximum satisfaction probability of directed controllers when the discount factor is sufficiently close to 1. So, in this section, we construct a supervisor via the optimal state-DCP value function and show that the supervisor achieves the maximum satisfaction probability and maximizes the number of the reachable states within the winning region.

 For the product SDES $D^\otimes$, we construct the following supervisor $SV^\ast : S^\otimes \to \Xi$. For each $s \in S^\otimes$,
\begin{align}
    \label{SV_ast}
    &SV^\ast(s) =\nonumber \\
    &\left\{
    \begin{aligned}
    & \bigcup \{ \xi_e \in \Xi^\otimes_{dir}(s) ; Q^\ast (s, \xi_e) = 0 \} & & \text{if } s \in W, \\
    & \xi_{e^\ast} & & \text{otherwise}, \\
    \end{aligned}
    \right.
\end{align}
where $\xi_{e^\ast} \in \argmax_{\xi_e \in \Xi^\otimes_{dir} (s)} Q^\ast(s,\xi_e)$. Note that $W$ is determined as $\{ s \in S^\otimes \;|\; V^\ast(s)=0 \}$ by (\ref{W}) and 1) of Lemma \ref{lemma_connect_V_Pr}.

We will show that the supervisor $SV^\ast$ achieves the maximum satisfaction probability of directed controllers when $\gamma$ is sufficiently close to 1. 

\begin{lemma}
 Given a product SDES $D^\otimes$ associated with an LTL formula $\varphi$ and the reward function $\mathcal{R}$ defined as (\ref{def_reward}), for any state $s \in S^\otimes$, the supervisor $SV^\ast$ defined as (\ref{SV_ast}) satisfies
\begin{align}
&Pr^{D^\otimes}_{SV^{\ast}}(s \models \varphi_B \land \textbf{F} W) = Pr^{D^\otimes}_{SV^{\ast}}(s \models \textbf{F} W). \label{GnAcc_FWinf_FWinf_SVast}
\end{align}
\end{lemma}

\begin{IEEEproof}
 It is sufficient to show that, for any $s \in S^\otimes$ and any $\rho  \in InfPath^{D^\otimes}_{SV^{d\ast}}(s)$, $\rho \models \textbf{F}W$ implies $\rho \models \varphi_B \land \textbf{F}W$. We fix a state $s \in S^\otimes$ and a path $\rho \in InfPath^{D^\otimes}_{SV^\ast}(s)$ arbitrarily. Suppose that $\rho \models \textbf{F} W$. Then, by Theorem \ref{thm_SVdast_winning} and 1) and 3) of Lemma \ref{lem_thm1}, we have $\rho \models \varphi_B$. Thus, for any $s \in S^\otimes$, (\ref{GnAcc_FWinf_FWinf_SVast}) holds.
\end{IEEEproof}

\begin{lemma}
\label{lemma_phiB_FWast}
Given a product SDES $D^\otimes$ associated with an LTL formula $\varphi$, for any state $s \in S^\otimes$, the supervisor $SV^\ast$ defined as (\ref{SV_ast}) satisfies.
\begin{align}
\label{PrSVast_PrSVWinf}
Pr^{D^\otimes}_{SV^{\ast}}(s \models \varphi_B) = Pr^{D^\otimes}_{SV^{\ast}}(s \models \textbf{F}W).
\end{align}
\end{lemma}

\begin{IEEEproof}
For any $s \in S^\otimes$, by (\ref{GnAcc_FWinf_FWinf_SVast}), we have
\begin{align*}
    &Pr^\otimes_{SV^{\ast}}(s \models \varphi_B) \\
    &= Pr^\otimes_{SV^{\ast}}(s \models \varphi_B \land (\textbf{F}W \lor \textbf{G}\neg W) ) \\
    &= Pr^{D^\otimes}_{SV^{\ast}}(s \models \textbf{F}W) + Pr^{D^\otimes}_{SV^{\ast}}(s \models \varphi_B \land \textbf{G}\neg W).
\end{align*}
By (\ref{SV_ast}), there exists an optimal directed controller $SV^{d\ast}$ such that $SV^\ast(s) = SV^{d\ast}(s)$ holds for any $s \not \in W$. Thus, (\ref{PrSVast_PrSVWinf}) holds by (\ref{GnAcc_GnWinf_0}).
\end{IEEEproof}

\begin{lemma}
\label{lemma_Prsvd_Prsv}
Given a product SDES $D^\otimes$ associated with an LTL formula $\varphi$ and the reward function $\mathcal{R}$ defined as (\ref{def_reward}), there exists an optimal directed controller $SV^{d\ast}$ such that, for any $s \in S^\otimes$,
\begin{align}
    \label{PrSVast_PrSVdast}
    Pr_{SV^\ast}^{D^\otimes}(s \models \varphi_B) = Pr_{SV^{d\ast}}^{D^\otimes}(s \models \varphi_B),
\end{align}
where the supervisor $SV^\ast$ is defined as (\ref{SV_ast}).
\end{lemma}

\begin{IEEEproof}
    By (\ref{SV_ast}), there exists an optimal directed controller $SV^{d\ast}$ such that $SV^{d\ast}(s) = SV^\ast(s)$ holds for any $s \not \in W$ and thus the following equation holds.
    \[
        Pr^{D^{\otimes}}_{SV^{d\ast}}(s \models \textbf{F}W)
        = Pr^{D^\otimes}_{SV^{\ast}}(s \models \textbf{F}W).
    \]
    Moreover, for any $s \in W$, clearly we have 
    \[
        Pr^{D^{\otimes}}_{SV^{d\ast}}(s \models \textbf{F}W)
        = Pr^{D^\otimes}_{SV^{\ast}}(s \models \textbf{F}W)
        = 1.
    \]
    Hence, by Proposition \ref{prop_phiB_FW_direct} and Lemma \ref{lemma_phiB_FWast}, (\ref{PrSVast_PrSVdast}) holds for any $s \in S^\otimes$.
\end{IEEEproof}

\begin{theorem}
Given a product SDES $D^\otimes$ associated with an LTL formula $\varphi$ and the reward function $\mathcal{R}$ defined as (\ref{def_reward}), there exists $\gamma' \in (0,1)$ such that, for any $\gamma > \gamma'$ and any $s \in S^\otimes$, the following equation holds.
\begin{align}
    \label{SVast_Pr_eq}
    Pr^{D^\otimes}_{SV^\ast}(s \models \varphi_B) = Pr^{D^\otimes}_{max^d}(s \models \varphi_B),
\end{align}
where the supervisor $SV^\ast$ is defined as (\ref{SV_ast}).
\label{thm_SVast_Pr}
\end{theorem}

\begin{IEEEproof}
    This immediately follows from Theorem \ref{thm_directed_main} and Lemma \ref{lemma_Prsvd_Prsv}. 
\end{IEEEproof}

By Theorem \ref{thm_SVast_Pr}, $SV^\ast$ defined as (\ref{SV_ast}) accomplishes the maximum satisfaction probability of directed controllers by taking the discount factor $\gamma$ sufficiently close to 1.

We define the notion of maximal permissiveness. 

\begin{definition}[Maximal permissiveness]
We say that a supervisor $SV \in \mathcal{SV}^{D^\otimes}_{sure}$ is maximally permissive if, for any supervisor $\widehat{SV} \in \mathcal{SV}^{D^\otimes}_{sure}$, the following inequality holds.
\begin{align}
    |Re_{SV}(S^\otimes)| \geq |Re_{\widehat{SV}}(S^\otimes)|.
\end{align}
\end{definition}

\begin{theorem}
\label{thm_maxperm}
Given a product SDES $D^\otimes$ associated with an LTL formula $\varphi$, the reward function $\mathcal{R}$ defined as (\ref{def_reward}), and the supervisor $SV^\ast$ defined as (\ref{SV_ast}), for any state $s \in W $, we have $SV^\ast \in \mathcal{SV}^{D^\otimes}_{sure}(s)$ and  $|Re_{SV^\ast}(S^\otimes;s)| \geq |Re_{\widehat{SV}}(S^\otimes;s)|$ for any $\widehat{SV} \in \mathcal{SV}^{D^\otimes}_{sure}(s)$.
\end{theorem}

\begin{IEEEproof}
For any $s \in W$, by Lemma \ref{lemma_phiB_FWast}, we have $Pr_{SV^\ast}^{D^\otimes}(s \models \varphi_B) = 1$. Thus, by 1) and 2) of Lemma \ref{lem_thm1}, $SV^\ast$ belongs to $\mathcal{SV}_{sure}^{D^\otimes}(s)$.
Suppose that there exists a supervisor $\widehat{SV} \in \mathcal{SV}^{D^\otimes}_{sure}(s)$ such that $|Re_{\widehat{SV}}(S^\otimes;s)| > |Re_{SV^\ast}(S^\otimes;s)|$. Then, there exist a state $\hat{s} \in W$ and a controllable event $\hat{e} \notin SV^\ast(\hat{s})$ such that $\hat{e} \in \widehat{SV}(\hat{s})$. By 2) and 3) of Lemma \ref{lem_thm1}, there exists a winning directed controller $\widehat{SV}^d$ with $\widehat{SV}^d (\hat{s}) = \xi_{\hat{e}}$ that is included by $\widehat{SV}$ but not by $SV^\ast$. Moreover, by 1) of Lemma \ref{lemma_connect_V_Pr}, $Q_{\widehat{SV}^d} (\hat{s},\xi_{\hat{e}}) = 0$. However, this contradicts (\ref{SV_ast}).
\end{IEEEproof}

\begin{remark}
 In synthesizing a supervisor that maximizes the size of the reachable set under the maximization of the satisfaction probability of $\varphi_B$, the directed control-based synthesis reduces computational complexity with respect to the size of the event set compared to dealing directly with control patterns. In detail, the exploration spaces for the directed control-based method and the straight forward one dealing with control patterns are proportion to $E_c$ and $2^{E_c}$, respectively.
\end{remark}

\begin{corollary}
\label{cor_maxpermissive}
Given a product SDES $D^\otimes$ associated with an LTL formula $\varphi$ and the reward function $\mathcal{R}$ defined as (\ref{def_reward}), if the initial state $s^{I \otimes} \in W$, then the supervisor $SV^\ast$ is maximally permissive.
\end{corollary}

\begin{IEEEproof}
 This immediately follows from Theorem \ref{thm_maxperm}.
\end{IEEEproof}

By Theorem \ref{thm_maxperm}, for any state $s$ in the winning region $W$, the number of the reachable states from $s$ is maximized while satisfying $\varphi_B$ surely under the supervisor $SV^\ast$ defined as (\ref{SV_ast}). Moreover, if $s^I \in W$ then $SV^\ast$ is maximally permissive.

\begin{remark}
By Corollary \ref{cor_maxpermissive}, it is sufficient to check whether the maximal value of the state-DCP function at the initial state is $0$ in order to confirm the maximal permissiveness of $SV^\ast$.
\end{remark}

In Example \ref{ex1}, the maximum state-DCP value of the initial state is 0. Thus, we construct the maximally permissive supervisor $SV^\ast$ as $SV^\ast((s_1, \{(x_0,0), (x_1,1)\}) = \{ b \}$ and $SV^\ast((s_2, \{(x_0,0)\})) = \{ a, c \}$.

\section{2-stage Reinforcement learning for Optimal Supervisors}
\label{2stageRL}

Practically, the dynamics of the SDES such as the transition probability is often unknown. Then, we use a reinforcement learning method to obtain an optimal directed controller.

For the product SDES associated with a given LTL formula, in order to learn the directed controllers that accomplish the maximum satisfaction probability, we decompose the problem of learning the desired directed controllers into the following two stages. The two-stage RL is partially based on Q-learning. 
\begin{LaTeXdescription}
    \item[\textnormal{Stage 1)}] We estimate the winning pair set $W_p$ by learning the state-DCP value function. Subsequently, we synthesize the winning directed controllers at any state in the winning region $W$ using the estimate of $W_p$. 
    \item[\textnormal{Stage 2)}] When the initial state does not belong to the winning region, we compute a directed controller that forces the paths on the product SDES to reach the winning region with the maximum probability by relearning the state value function for the states outside the winning region.
\end{LaTeXdescription}
Note that the decomposition is based on Proposition \ref{prop_phiB_FW_direct}. In other words, learning of the desired supervisor is decomposed into 1) learning $W_p$ (the behavior in $W$) and (2) learning a directed controller that maximizes the probability of reaching $W$ (the behavior outside $W$).

\subsection{Learning of Winning pair set and Winning directed controllers}
For Stage 1), we propose Algorithm \ref{ql2} based on Q-learning so as to obtain the winning pair set.

For a product SDES $D^\otimes$, let $Q^k$ be a state-DCP value function learned up to the episode $k \in \mathbb{N}$ by Algorithm \ref{ql2}.

We call the following set of pairs of states and directed control patterns an \textit{estimated winning pair set} learned up to the episode $k \in \mathbb{N}_0$.
\begin{align}
    W^k_{p} = \{ (s,\xi_e) \;|\; s \in S^\otimes, \xi_e \in \Xi^\otimes_{dir}(s),  Q^k(s,\xi_e) = 0\}.
    \label{Wkp}
\end{align}

We call the following subset of $S^\otimes \setminus Acc^\otimes$ an \textit{estimated winning region} learned up to the episode $k \in \mathbb{N}_0$.
\begin{align}
    W^k = \{ s \in S^\otimes \;|\; \exists \xi_e \in \Xi^\otimes_{dir}(s) \text{ s.t. } Q^k(s,\xi_e) = 0\}.
    \label{Wk}
\end{align}
 
 \begin{algorithm}
 \caption{Learning of $W_p$ and $W$.}
 \label{ql2}
 \begin{algorithmic}[1]
 \renewcommand{\algorithmicrequire}{\textbf{Input:}}
 \renewcommand{\algorithmicensure}{\textbf{Output:}}
 \REQUIRE LTL formula $\varphi$ and SDES $D$.
 \ENSURE The state-DCP value function $Q^\infty$ such that, for any $s \in S^\otimes$ and $\xi_e \in \Xi^\otimes_{dir}(s)$, $Q^\infty(s,\xi_e) = 0$ if and only if $Q^\ast(s,\xi_e) = 0$.
 \STATE Convert $\varphi$ to d$K$cBA $det(B_{\varphi},K)$.
  \STATE Construct the product SDES  $D^{\otimes}$ of $D$ and $det(B_{\varphi},K)$.
  \STATE Initialize $Q^k(s,\xi_e)$ with 0 for any $s \in S^\otimes \setminus Acc^\otimes$ and any $\xi_e \in E^\otimes_{c,dir}(s)$, and $r_n$ for any $s \in Acc^\otimes$ and any $\xi_e \in E^\otimes_{c,dir}(s)$.
  \STATE Compute $W^0_p$ and $W^0$. 
  \STATE Pick up $s \in W^0$ at random and set $k=0$.
  \WHILE {$W^k_p$ does not converge}
  \WHILE {True}
  \STATE Choose $\xi_e$ from $\{ \xi_e \in \Xi^\otimes_{dir}(s) \;|\; Q^k(s,\xi_e) = 0 \}$ uniformly and at random.
  \STATE Observe an event $e_{oc} \in \xi_e$ and the next state $s'$.
  \IF{$s' \not\in W^k$.}
  \STATE Go to Line 14.
  \ENDIF
  \STATE $s \gets s'$.
  \ENDWHILE
  \IF{ $e_{oc}$ is uncontrollable.}
  \FOR{ all events $\xi_e \in \Xi^\otimes_{dir}(s)$ }
  \STATE \begin{align*} 
      Q^{k+1}& (s, \xi_e)\! \leftarrow\!  (1-\alpha)Q^k(s,\xi_e) + \alpha \{ \mathcal{R}(s') \\
      & + \Gamma(s') \max_{\xi_{e^\prime} \in \Xi^\otimes_{dir}(s')}Q^k(s',\xi_{e^\prime}) \}.
  \end{align*}
  \ENDFOR
  \ELSE
  \STATE \begin{align*} 
      Q^{k+1}& (s,\xi_e) \leftarrow (1 - \alpha) Q^k(s,\xi_e) + \alpha \{ \mathcal{R}(s') \\
      &+ \Gamma(s') \max_{\xi_{e^\prime} \in \Xi_{dir}^\otimes(s') }Q^k(s', \xi_{e^\prime}) \}.
  \end{align*}
  \ENDIF
  \STATE $k \gets k+1$.
  \STATE Compute $W^k_p$ and $W^k$ and pick up $s \in W^k$.
  \ENDWHILE
 \end{algorithmic}
\end{algorithm}

We now describe Algorithm \ref{ql2}. Recall that the state values of unsafe states and states in the winning region are $r_n$ and 0, respectively. Thus, at Line 3, we initialize the state-DCP value function with 0 and $r_n$ for the safe states and the unsafe states, respectively. $W^0$ and $W^0_p$ are initialized with $S^\otimes \setminus Acc^\otimes$ and $\{(s,\xi) \;|\; s \in S^\otimes \setminus Acc^\otimes, \xi \in \Xi^\otimes_{dir}(s) \}$, respectively.
At Line 6, Algorithm \ref{ql2} continues until the estimated winning pair set converges. At Line 7, for a current state $s \in W^k$, we choose a directed control pattern $\xi_e$ from $\{ \xi_e \in \Xi^\otimes_{dir}(s) \;|\; Q^k(s,\xi_e) = 0\}$. An event $e_{oc} \in \xi_e$ occurs and observe the next state $s'$. If $s' \not \in W^k$, we terminate the exploration in the episode.
In Line 15, if $e_{oc}$ is uncontrollable, i.e., the state leaves from the current estimated winning region by an uncontrollable event, we update the state-DCP values for all directed control patterns. Other than that, we update the state-DCP value of $s$ and $\xi_e$ in Line 20. From Line 22 to 24, we update $W^k$ and $W^k_p$, pick up $s$ from $W^k$, and continue the learning.

 
We will show that Algorithm \ref{ql2} estimates the winning pair set and synthesize all winning directed controllers for the winning region.
\begin{lemma}
 Given a product SDES $D^\otimes$ associated with an LTL formula $\varphi$ and the reward function $\mathcal{R}$ defined as (\ref{def_reward}), when conducting Algorithm \ref{ql2}, for any state $s \in S^\otimes \setminus Acc^\otimes$ and any $\xi_e \in \Xi^\otimes_{dir}(s)$, if there exists $k' \in \mathbb{N}_0$ such that $Q^{k'}(s,\xi_e) < 0$, then, for any $k \geq k'$, we have $Q^k(s,\xi_e) < 0$.
 \label{lem_Qk1_Qk}
\end{lemma}

\begin{IEEEproof}
 We fix a state $s \in S^\otimes$ and $\xi_e \in \Xi^\otimes_{dir}(s)$ arbitrarily. Suppose that there exists $k' \in \mathbb{N}_0$ such that $Q^{k'}(s,\xi_e) < 0$.
 Note that the reward $\mathcal{R}(s)$ is 0 or negative. Thus, we have $Q^k(s,\xi_e) < 0$ for any $k \geq k'$.
\end{IEEEproof}
Lemma \ref{lem_Qk1_Qk} implies that $W^k_p$ and $W^k$ are monotonically decreasing with respect to $k$.
\begin{align}
     W^{k+1}_p \subseteq W^k_p.
     \label{rel_Wk1p_Wkp}
 \end{align}
 \begin{align}
     W^{k+1} \subseteq W^k.
     \label{rel_Wk1_Wk}
 \end{align}
Hence, there exist $W^\infty_p = \lim_{k \to \infty} W^k_p$ and $W^\infty = \lim_{k \to \infty} W^k$.

\begin{lemma}
 Given a product SDES $D^\otimes$ associated with an LTL formula $\varphi$ and the reward function $\mathcal{R}$ defined as (\ref{def_reward}), for any $s \in S^\otimes \setminus Acc^\otimes$ and any $\xi_e \in \Xi^\otimes_{dir}(s)$, if there exists $k \in \mathbb{N}$ such that  $Q^k(s,\xi_e) = 0$ and $Q^{k+1}(s,\xi_e) < 0$, then there exists a state $s' \in S^\otimes \setminus W^k$ such that $P^\otimes(s'|s,\xi_e) > 0$.
 \label{lem_Qk_outtrans}
\end{lemma}
\begin{IEEEproof}
We fix a state $s \in S^\otimes$ and $\xi_e \in \Xi^\otimes_{dir}(s)$ arbitrarily. Suppose that there exists $k \in \mathbb{N}_0$ such that $Q^k(s, \xi_e) = 0$ and $Q^{k+1}(s, \xi_e) < 0$. Then, by the update rule of the state-DCP value function in Algorithm \ref{ql2}, there exists an event $e' \in \xi_e$ such that it triggers an outgoing transition from $s$ to $S^\otimes \setminus W^k$ with a positive probability. Thus, there exists a state $s' \in S^\otimes \setminus W^k$ such that $P^\otimes(s'|s,\xi_e) > 0$.
\end{IEEEproof}

\begin{prop}
 Given a product SDES $D^\otimes$ associated with an LTL formula $\varphi$ and the reward function $\mathcal{R}$ defined as (\ref{def_reward}), when conducting Algorithm \ref{ql2}, for any state $s \in S^\otimes$, any $\xi_e \in \Xi^\otimes_{dir}(s)$, and any $k \in \mathbb{N}_0$, if $Q^k(s,\xi_e) < 0$ holds, then we have $Pr^{D^\otimes}_{max^d}(s,\xi_e \models \varphi_B) < 1$.
 \label{prop_Wk_Pr}
\end{prop}

\begin{IEEEproof}
 We prove by induction with respect to $k \in \mathbb{N}_0$. Consider the case where $k = 0$. Then, for any $s \in S^\otimes$ and any $\xi_e \in \Xi^\otimes_{dir}(s)$, if $Q^0(s,\xi_e) < 0$ holds, then we have $s \in Acc^\otimes$ and thus $Pr_{max^d}^{D^\otimes}(s, \xi_e \models \varphi_B) < 1$. Let $k \in \mathbb{N}_0$ be given. For any $s \in S^\otimes \setminus Acc^\otimes$ and any $\xi_e \in \Xi^\otimes_{dir}(s)$, suppose that $Q^k(s,\xi_e)<0$ implies $Pr_{max^d}^{D^\otimes}(s, \xi_e \models \varphi_B) < 1$. Then, for $k+1$, any $s \in S^\otimes \setminus Acc^\otimes$, and any $\xi_e \in \Xi^\otimes_{dir}(s)$, suppose that $Q^{k+1}(s, \xi_e) < 0$. Consider the two cases 1) $Q^k(s,\xi_e)<0$ and 2) $Q^k(s,\xi_e)=0$. For the case 1), we immediately have $Pr_{max^d}^{D^\otimes}(s, \xi_e \models \varphi_B) < 1$ by the induction hypothesis. For the case 2), by Lemma \ref{lem_Qk_outtrans}, there exists a state $s' \not\in W^k$ such that $P^\otimes(s'|s,\xi_e)>0$.
 Thus, we have the same result as the case 1) by applying the induction hypothesis to $s'$ and all $\xi_{e'} \in \Xi^\otimes_{dir}(s')$. Thus, Proposition \ref{prop_Wk_Pr} is proved.
\end{IEEEproof}

Proposition \ref{prop_Wk_Pr} implies that, for the winning region $W$, the winning pair set $W_p$, and any $k \in \mathbb{N}_0$, the following relations hold.
\begin{align}
     W_p \subseteq W^k_p. \label{rel_Wp_Wkp} \\
     W \subseteq W^k.
     \label{rel_W_Wk}
 \end{align}
 Due to (\ref{rel_Wk1p_Wkp}), (\ref{rel_Wk1_Wk}), (\ref{rel_Wp_Wkp}), and (\ref{rel_W_Wk}), for each $k \in \mathbb{N}_0$ and each $s \in W^k$, we pick up a state from $W^k$ at the first step of each episode, restrict the candidates of directed control pattern to be chosen to $\{ \xi_e \in \Xi^\otimes \;|\; Q^k(s,\xi_e) = 0 \}$, and terminate a current episode when the state leaves from $W^k$ in Algorithm \ref{ql2}.


\begin{assumption}
 Given a product SDES $D^\otimes$ associated with an LTL formula $\varphi$, when conducting Algorithm \ref{ql2}, the following conditions hold.
 \begin{enumerate}
     \item The learning ratio $\alpha \in (0,1)$ is a constant.
     \item For each $s \in W^\infty$, $s$ is observed infinitely often with probability 1.
 \end{enumerate}
 \label{assum_W_s_infty}
\end{assumption}

To show that $W^\infty_p = W_p$ holds with probability 1, we define the following supervisor $SV^\infty$ constructed via $W^\infty_p$. For any $s \in S^\otimes$,

\begin{align}
    SV^\infty(s) = 
    \left\{
    \begin{aligned}
      & \bigcup_{ \{ \xi \;|\; (s,\xi) \in W^\infty_p \} } \xi & &\text{if} \ s \in W^\infty , \\
      & E^\otimes(s) & &\text{otherwise}.
    \end{aligned}
    \right.
    \label{SVinf}
\end{align}

\begin{lemma}
 Given a product SDES $D^\otimes$ associated with an LTL formula $\varphi$ and the reward function $\mathcal{R}$ defined as (\ref{def_reward}), if $Re_{SV^\infty}(S^\otimes;s) \subseteq W^\infty$ holds for any $s \in W^\infty$, then $W^\infty_p \subseteq W_p$ holds.
 \label{lem_Re_Winf_Wpinf_Wp}
\end{lemma}
\begin{IEEEproof}
 Let $SV^\infty$ be the supervisor defined as (\ref{SVinf}). Suppose that $Re_{SV^\infty}(S^\otimes;s) \subseteq W^\infty$ holds for any $s \in W^\infty$. Then, we have $s' \not \in Acc^\otimes$ for any $s' \in Re_{SV^\infty}(S^\otimes;s)$ since $W^\infty \cap Acc^\otimes = \emptyset$. This implies together with 1) and 3) of Lemma \ref{lem_thm1} that, for any directed control pattern $\xi \subseteq SV^\infty(s)$, we have $(s,\xi) \in W_p$. Thus, $W^\infty_p \subseteq W_p$ holds.
\end{IEEEproof}

\begin{prop}
 Given a product SDES $D^\otimes$ associated with an LTL formula $\varphi$ and the reward function $\mathcal{R}$ defined as (\ref{def_reward}), we have $W^\infty_p = W_p$ with probability 1 under Assumption \ref{assum_W_s_infty}.
 \label{prop_Wp_W}
\end{prop}

\begin{IEEEproof}
Let $SV^\infty$ be the supervisor defined as (\ref{SVinf}). For simplicity, we abbreviate ``with probability 1" as ``w.p.1". By (\ref{rel_Wp_Wkp}) and Lemma \ref{lem_Re_Winf_Wpinf_Wp}, it is sufficient to show that, for any $s \in W^\infty$, $Re_{SV^\infty}(S^\otimes;s) \subseteq W^\infty$ holds w.p.1. Suppose that, there exists $s' \in Re_{SV^\infty}(S^\otimes;s)$ such that $s' \not \in W^\infty$. For any $\{s_i\}_{i=0}^m \subseteq Re_{SV^\infty}(S^\otimes;s)$ such that $P^\otimes(s_{i+1} | s_i, SV^\infty(s_i)) > 0$ for each $i = 0,\ldots,m-1$, $s_0=s$, and $s_m = s'$, there exists $n  \leq m$ such that $s_{n-1} \in W^\infty$ and $s_n \not \in W^\infty$. Thus, there exists $k \in \mathbb{N}$ such that $s_n \in W^{k-1}$ and $s_n \not\in W^k$. By 2) of Assumption \ref{assum_W_s_infty}, $s_{n-1}$ is visited infinitely often w.p.1. Thus, w.p.1, for any $e \in SV^\infty(s_{n-1})$, the transition $(s_{n-1},e,s_n)$ occurs infinitely often. Hence, w.p.1, there exists $l > k$ such that $(s_{n-1}, e, s_n)$ is observed at the step $l$. Thus, by 1) of Assumption \ref{assum_W_s_infty}, w.p.1, for any $\xi \subseteq SV^\infty(s_{n-1})$, if $e \in \xi$ then $Q^l(s_{n-1},\xi) < 0$. Thus, this contradicts that $\xi \subseteq SV^\infty(s_{n-1})$, which implies that $Re_{SV^\infty}(S^\otimes;s) \subseteq W^\infty$ holds w.p.1.
\end{IEEEproof}

Proposition \ref{prop_Wp_W} implies $W^\infty = W$ with probability 1. Moreover, by 1) of Lemma \ref{lemma_connect_V_Pr}, we have that, for any $s \in S^\otimes$ and any $\xi \in \Xi^\otimes_{dir}(s)$, $Q^\infty(s,\xi) = 0$ if and only if $Q^\ast(s,\xi)=0$.

\subsection{Learning of directed controllers with maximum satisfaction probability}
After conducting Algorithm \ref{ql2}, if the initial state does not belong to $W^\infty$, we have to compute a directed controller that forces the paths on the controlled product SDES to reach the winning region with the maximum probability.
For Stage 2), we propose Algorithm \ref{ql1} based on Q-learning\footnote{When the initial state is in the winning region, we do not have to conduct Algorithm \ref{ql1}.}.

We conduct Algorithm \ref{ql1} on the same product SDES given to Algorithm \ref{ql2} with the state-DCP value function learned by it. At Line 4, if the current state is in $W^\infty$, we terminate the episode. In other words, we compute the state values of states outside the estimated winning region during Algorithm \ref{ql1}. Note that, in practice, we terminate an episode in Algorithm \ref{ql1} after entering $Acc^\otimes$. 

We denote by $Q^\infty$ the state-DCP value function obtained from Algorithm \ref{ql1}.
\begin{prop}
\label{prop_phiB_FWast_direct}
 Given a product SDES $D^\otimes$ associated with an LTL formula $\varphi$ and the reward function $\mathcal{R}$ defined as (\ref{def_reward}), for any state $s \in S^\otimes$ and any $\xi \in \Xi^\otimes_{dir}(s)$, we have $Q^\infty(s,\xi) = Q^\ast(s,\xi)$ with probability 1 under the assumption that $\sum_{k=0}^\infty \alpha_k = \infty$ and $\sum_{k=0}^\infty \alpha_k^2 < \infty$ hold in Algorithm \ref{ql1}.
\end{prop}

\begin{IEEEproof}
 Note that the correct optimal state value of any state in $W^\infty$ is 0 with probability 1 by Proposition \ref{prop_Wk_Pr}. The proof follows from Proposition 4.5 in \cite{BT1996}.
\end{IEEEproof}

By Propositions \ref{prop_Wp_W} and \ref{prop_phiB_FWast_direct}, using Algorithms \ref{ql2} and \ref{ql1}, we can obtain the same permissive supervisor $SV^\ast$ defined as (\ref{SV_ast}) by replacing $Q^\ast$ and $W$ with $Q^\infty$ and $W^\infty$, respectively, with probability 1.

\begin{algorithm}
 \caption{Q-learning of $SV^d$ maximizing the satisfaction probability of $\varphi_B$.}
 \label{ql1}
 \begin{algorithmic}[1]
 \renewcommand{\algorithmicrequire}{\textbf{Input:}}
 \renewcommand{\algorithmicensure}{\textbf{Output:}}
 \REQUIRE The product SDES $D^\otimes$ given to Algorithm \ref{ql2} and the state-DCP value function learned by Algorithm \ref{ql2}.
 \ENSURE The optimal state-DCP value function $Q^\ast$.
  \STATE Let $k = 0$.
  \WHILE {$Q^k$ does not converge}
  \STATE $s \leftarrow (s^{I},F^I)$.
  \FOR {$s \not \in W^\infty$}
  \STATE Choose $\xi_e \in \Xi_{dir}^\otimes(s) $. 
  \STATE An event $e_{oc} \in \xi_e$ occurs.
  \STATE Observe the next state $s^{\prime}$.
  \STATE Obtain the reward $\mathcal{R}(s^{\prime})$.
  \STATE
  \begin{align*}
      Q^{k+1}(s,\xi_e) \leftarrow & (1 - \alpha_k) Q^k(s,\xi_e) + \alpha_k \{ \mathcal{R}(s^{\prime})\\
      &+ \Gamma(s^{\prime}) \max_{\xi_{e^\prime} \in \Xi_{dir}^\otimes(s^\prime) }Q^k(s^{\prime},\xi_{e^\prime}) \}.
    \end{align*}
  \STATE $s \leftarrow s'$.
  \ENDFOR
  \STATE $k \gets k+1$.
  \ENDWHILE
 \end{algorithmic}
\end{algorithm}

\section{Example}
\label{EX}

\begin{figure}[htbp]
  	\centering
  	\subfigure[]{
  		\includegraphics[width = 3.5cm]{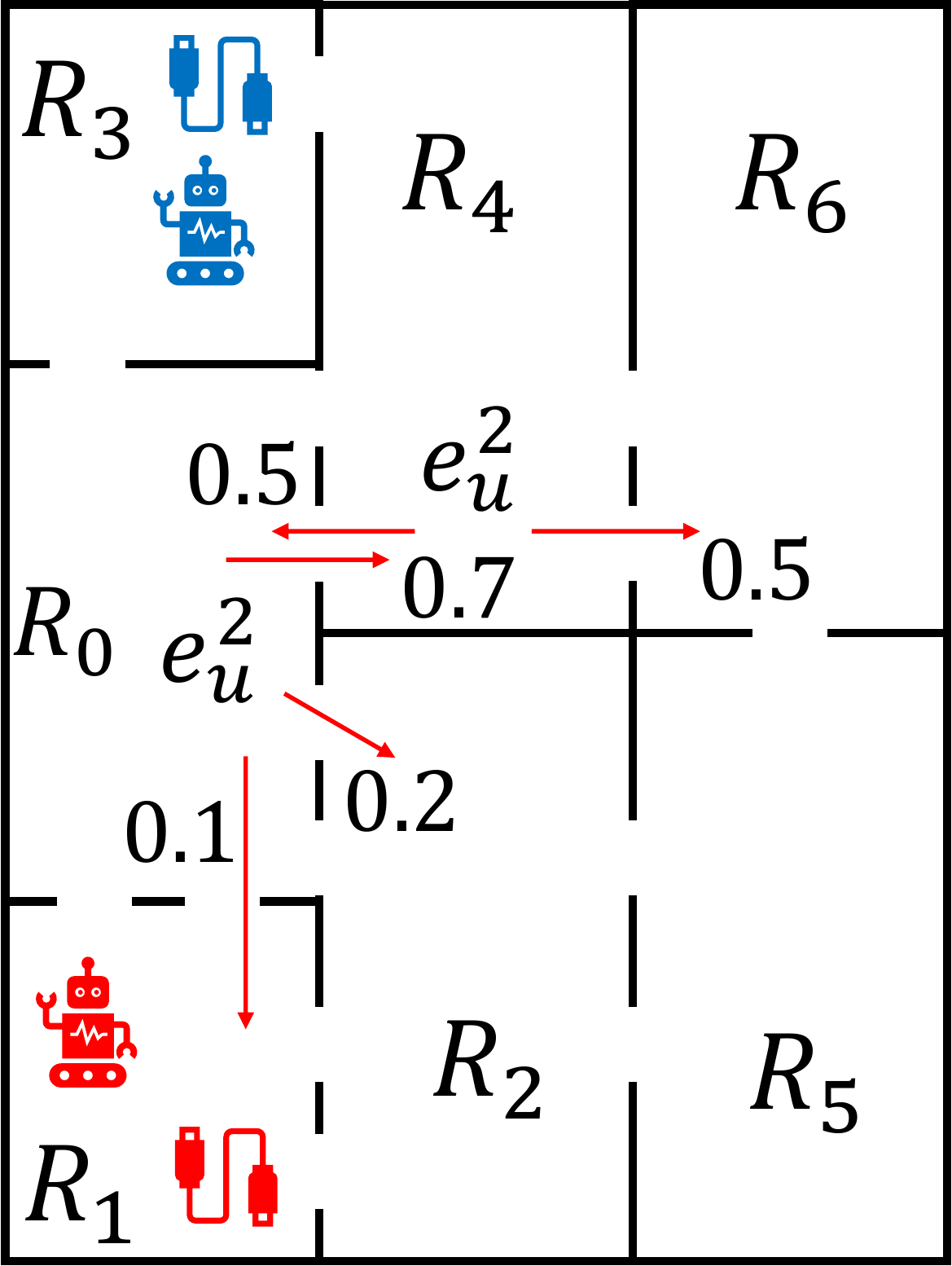}
        \label{ex:env}
  	}
  	\subfigure[]{
  		\includegraphics[width=3.6cm]{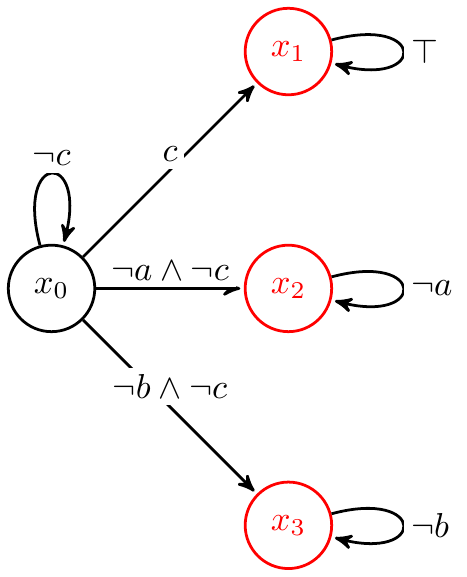}
  		\label{ex:automaton}
  	}
  	\caption{(a) An illustrative example of environment consisting of 7 rooms. Recharge rooms for the robot 1 (blue) and the robot 2 (red) are rooms $R_3$ and $R_1$, respectively. The uncontrollable events $e^2_{u}$ provokes unintended moves of the robot 2. the red arrows and their labels indicate the directions of triggered moves and their transition probabilities, respectively. (b) A cBA converted from $\varphi = \textbf{GF} a \land \textbf{GF} b \land \textbf{G} \neg c$. The accepting states are colored with red.}
  \end{figure}

We apply the proposed method to a motion planning problem with two robots. They move in an inner environment that consists of 7 rooms shown in Fig.\! \ref{ex:env}. The state space of the SDES is $S = \{ (s_1, s_2) ; s_j \in \{R_i\}_{i=1}^7,\ j=1,2 \}$, where $s_1$ and $s_2$ represent the rooms in which the respective robot is and $R_i$ with $i = 1,\ldots,7$ represents a room.
The two robots survey on the environment. Let $E = \{ e^1_{0}, e^1_{1}, e^1_{2}. e^1_{3}, e^1_{4}, e^1_{6}, e^2_{0}, e^2_{1}, e^2_{2}, e^2_{4}, e^2_{5}, e^2_{6}, e^2_{u} \}$ be the set of events, where each event indicates a movement of a robot.
The event $e^k_{i}$ represents that the robot $k$ tries to move to the next room $R_i$ from the current room. For example, when the event $e^1_{1}$ occurs at $(s_1, s_2) = (R_2,R_3)$, the robot 1 moves to the next room $R_1$ from $R_2$ while the robot 2 stays at the current room. The event $e^2_{u}$ is an uncontrollable and it can occur when the robot 2 is in $R_0$ or $R_4$, which means the event $e^2_{u}$ trigger unintended moves of the robot 2. The transitions from $R_0$ to $R_2$, $R_3$, and $R_4$ by the event $e^2_{u}$ occur with probabilities 0.1, 0.2, and 0.7, respectively. 
The both transitions from $R_4$ to $R_0$ and $R_6$ by $e^2_u$ occur with probability 0.5.
Similarly, the controllable event $e^1_{4}$ triggers the probabilistic transition when the robot 1 is in $R_3$. Associated with $e^1_{4}$, the transitions from $R_3$ to $R_0$ and $R_4$ occur with the probability 0.5. The other transitions are deterministic, that is, the robots move to the intended directions. We require both robots to return their recharge rooms (rooms 1 and 3) infinitely often while avoiding being in the same room at the same time. The specification is represented formally as $\textbf{GF}\text{r1} \land \textbf{GF}\text{r2} \land \textbf{G}\neg\text{u}$, where r1 means that the robot 1 is in $R_3$, r2 means that the robot 2 is in $R_1$, and u means that both robots are in the same room. A cBA converted from $\varphi$ is shown in Fig. \ref{ex:automaton}. We use the d$K$cBA obtained from the cBA with $K = 10$.

We set $\gamma = 0.9999$, $\gamma_{acc} = 0.9$, $r_n=-1$, and $s^I = (R_0,R_3)$. We conduct the proposed 2-stage reinforcement learning method. We train directed controllers with 4000 and 200000 episodes for Algorithms \ref{ql2} and \ref{ql1}, respectively. We iterate $T_{epi}=5000$ steps per episode and repeat 100 learning sessions. For the stage 1, when sampling an initial state from the estimated winning region, we use the probability distribution $p^k : W^k \to [0,1]$ at each step $k \in \mathbb{N}_0$ defined as 
\begin{align*}
    p^k(s) =
    \frac{ \left(\sum_{t=0}^{k} \mathds{1}_{s_t}(s) \right)^{-1} }{ \sum_{s \in W^k} \left(\sum_{t=0}^{k} \mathds{1}_{s_t}(s) \right)^{-1} },
\end{align*}
where $s_t$ is the state at the step $t$. Intuitively, for any $k \in \mathbb{N}_0$ and any $s \in W^k$, the smaller the times of visiting $s$ are in the learning, the greater the probability of sampling $s$ is.

\subsection*{Results for Stage 1 (Algorithm \ref{ql2})}
\begin{figure}[tbp]
 \centering
     \includegraphics[ width=7cm]{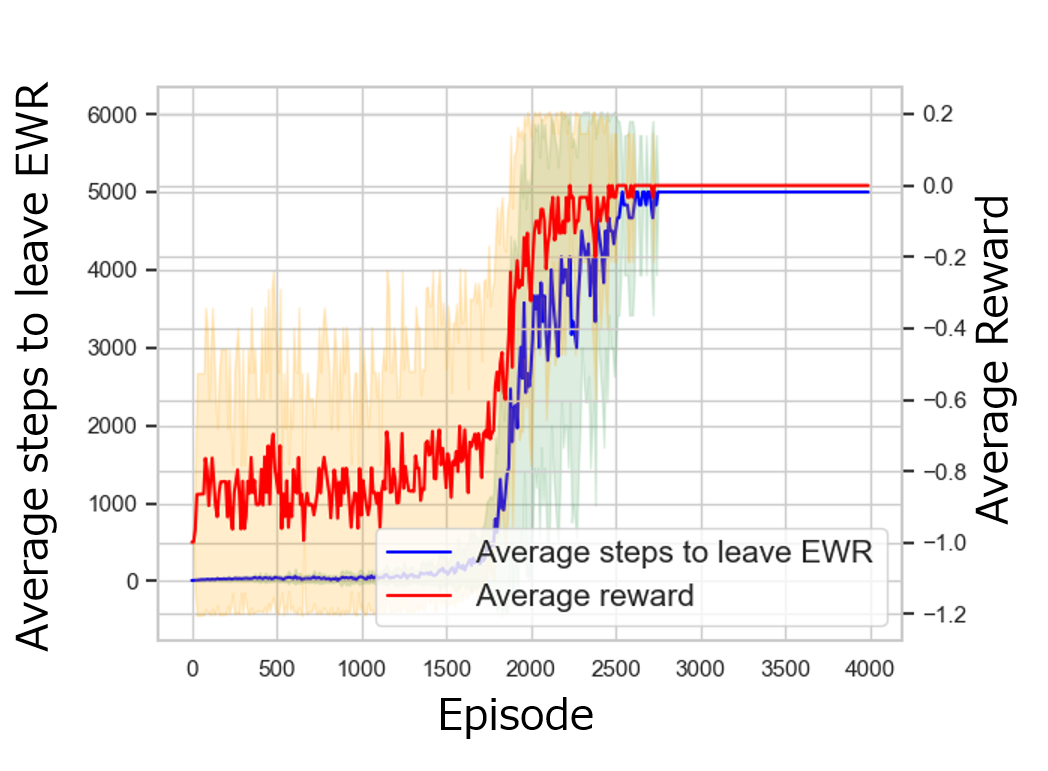}
    \caption{Mean of the average reward plotted by the red line and the average steps to leave the estimated winning region (EWR) plotted by the blue line in each episode for 100 learning sessions obtained by Algorithm \ref{ql2}. The orange and green shaded areas represent the ranges of the standard deviations.} 
    \label{average_reward_steps}
\end{figure}

\begin{figure}[tbp]
 \centering
   \includegraphics[width=6.5cm]{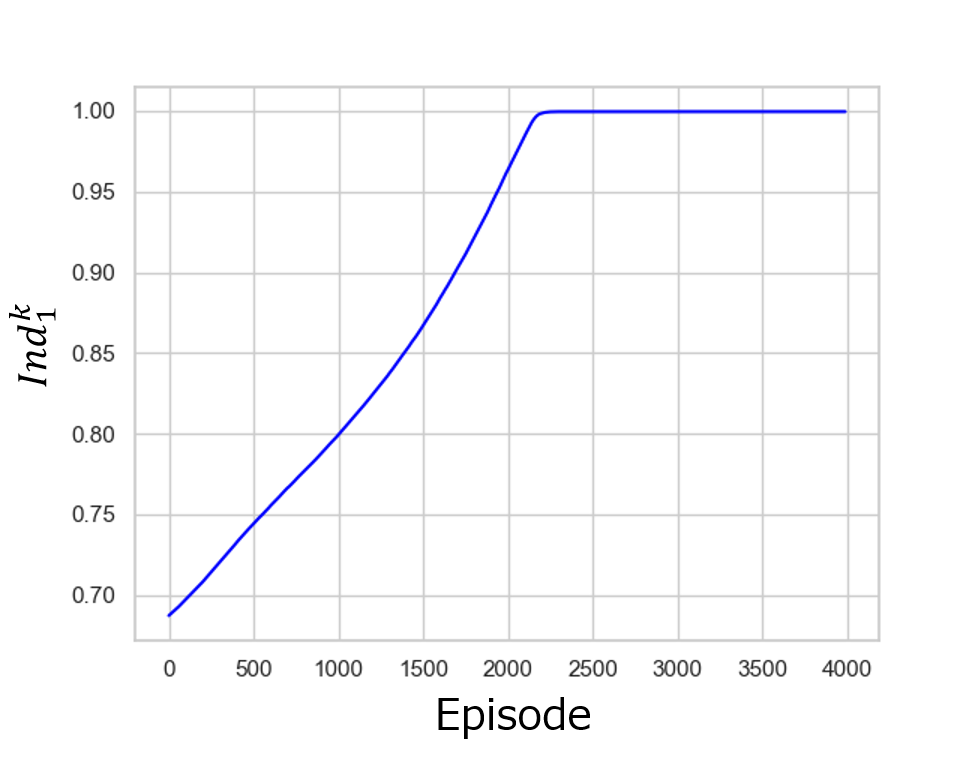}
 \caption{Average of $Ind^k_1$ of true $SV^\ast$ and a learned supervisor up to episode $k$ during Algorithm \ref{ql2}. }
 \label{coincide}
\end{figure}

Shown in Fig.\ \ref{average_reward_steps} are the average reward and the average steps to leave the estimated winning region per episode obtained by Algorithm \ref{ql2}. They converge to 0 and 5000, respectively, by 2500 episodes. Moreover, we observe a lot of episodes where the obtained reward is 0 before the 2500-th episode. This is because the robots tend to move only within the estimated winning region during Algorithm \ref{ql2}, that is, the robots try to avoid reaching an unsafe state. 
To show that the learned supervisor maximizes the reachable set within the winning region, we introduce the following index $Ind^k_1$ for each episode $k$.
\[
Ind^k_1 = \frac{\sum_{s \in W} |SV^\ast(s) \setminus E^\otimes_{uc}(s)|}{\sum_{s \in W} |SV^k(s) \setminus E^\otimes_{uc}(s)|},
\]
where $W$ is the winning region of the product SDES and $SV^k$ is a supervisor constructed from $Q^{k}$ in the same way as (\ref{SV_ast}). When $Ind^k_1$ becomes 1, the learned supervisor maximizes the reachable set within the winning region because, for each state in the winning region, all control patterns at the state of both the supervisor $SV^\ast$ and the leaned one are exactly the same.
Shown in Fig.\ \ref{coincide} is the average of $Ind^k_1$ and it converges to 1. This implies that the supervisor obtained by Algorithm \ref{ql2} maximizes the reachable set within the winning region.

\subsection*{Results for Stage 2 (Algorithm \ref{ql1})}
We introduce an index $Ind^k_2$ to evaluate whether the obtained supervisor maximizes the probability of reaching the estimated winning region. Let $Re_{SV^\ast}^{0<p<1}(S^\otimes) = Re_{SV^\ast}(S^\otimes) \setminus ( Acc^\otimes \cup W^\infty)$. The index $Ind^k_2$ is defined as
\begin{align*}
 Ind^k_2  = \max_{SV^\ast} \frac{ \sum_{s \in Re_{SV^\ast}^{0<p<1}(S^\otimes)} | SV^\ast(s) \cap SV^k(s) \setminus E^\otimes_{uc}(s)| } { | Re_{SV^\ast}^{0<p<1}(S^\otimes) | }.
\end{align*}
Intuitively, $Ind^k_2$ represents the coincidence ratio between the supervisor $SV^\ast$ and the supervisor learned up to the episode $k$ in the states outside the estimated winning region and the accepting set.
Shown in Table \ref{match_ratio} is the frequency of each value of $Ind^k_2$ for the last episode of each learning session. Note that $Ind^k_2 \in \{ 0,1/9,\ldots,8/9,1\}$. $Ind^k_2$ is $7/9$ or more in all sessions and moreover $Ind^k_2 = 1$ in 52 sessions. This implies that the supervisor obtained by Algorithm \ref{ql1} approaches a supervisor that achieves the maximum satisfaction probability.

\begin{table}[htbp]
  \caption{The frequency of each value of $Ind^k_2$ for the last episodes.}
  \label{match_ratio}
  \centering
    \begin{tabular}{ l l } \hline
       Value of $Ind^k_2$ & Frequency \\ \hline 
       $Ind^k_2 \in \{ 0, \ldots, 2/3 \}$ & 0 \\
       $Ind^k_2 = 7/9$ & 10 \\
       $Ind^k_2 = 8/9$ & 38 \\
       $Ind^k_2 = 1$ & \textbf{52} \\ \hline
    \end{tabular}
\end{table}

\section{Conclusion}
\label{CON}
This paper proposed a novel value-based synthesis of supervisors for stochastic discrete event systems (SDES).  For a given LTL formula and a design parameter $K$, we constructed a universal $K$-co-B{\"u}chi automaton and synthesized optimal directed controllers such that the probability of the acceptance of controlled behaviors by the automaton is maximized. Then, we considered the synthesis of an optimal supervisor and showed that it maximizes the reachability set under the maximization of the probability of the acceptance. Moreover, we considered the case where the system is unknown and proposed a 2-stage reinforcement learning based method for an efficient synthesis of the desired supervisor and showed that it synthesizes an optimal supervisor with probability 1 under some mild conditions. Future works are to extend the proposed method to uncountable state spaces, to combine with safe learning methods, and to analyze the convergence rate of the proposed method.

\bibliographystyle{ieeetr}
\bibliography{reference}

\begin{IEEEbiography}{Ryohei Oura}
    R.~Oura received B.E. degree in 2020 from Osaka University, Osaka, Japan. Currently, he is a student at the Graduate School of Engineering Science in Osaka University. His research interests include formal methods and machine learning.
\end{IEEEbiography}
\begin{IEEEbiography}{Toshimitsu Ushio}
       T.~Ushio (S'84--M'85) received the B.E.,
       M.E., and Ph.D.~degrees in 1980, 1982, and 1985,
       respectively, from Kobe University, Kobe, Japan.
       He joined Osaka University as an Associate Professor
       in 1994, and is currently a Professor. His
       research interests include control and analysis of
       discrete event systems and hybrid systems.
       Dr.~Ushio is a member of IEICE, SICE, and ISCIE.
\end{IEEEbiography}
\begin{IEEEbiography}{Ami Sakakibara}
A.~Sakakibara received the B.E., M.E., and Ph.D.~degrees in 2015, 2017, and 2020, respectively, from Osaka University, Japan. 
She was a Research Fellow of the Japan Society for the Promotion of Science from 2019 to 2020. 
Currently, she is a postdoctoral researcher at KTH Royal Institute of Technology, Sweden. 
Her research interests include formal methods, supervisory control and game theory.
Dr.~Sakakibara is a member of IEICE.
\end{IEEEbiography}

\end{document}